\documentclass[prc,twocolumn,showpacs,preprintnumbers,amsmath,amssymb]{revtex4-1}
\usepackage[dvips]{graphicx}
\usepackage{dcolumn}
\usepackage{bm}
\usepackage{color}

\begin{document}
\title{Extension of time-dependent Hartree-Fock-Bogoliubov equations}

\author{Mitsuru Tohyama}
\affiliation{Kyorin University School of Medicine, Mitaka, Tokyo
  181-8611, Japan} 
\author{Peter Schuck}
\affiliation{Institut de Physique Nucl$\acute{e}$aire, IN2P3-CNRS,
Universit$\acute{e}$ Paris-Sud, F-91406 Orsay Cedex, France}
\affiliation{Laboratoire de Physique et de Mod\'elisation des Milieux Condens\'es, CNRS 
 et, Universit\'e Joseph Fourier, 25 Av. des Martyrs, BP 166, F-38042 
 Grenoble Cedex 9, France} 
\begin{abstract}
An extension of the time-dependent Hartree-Fock-Bogoliubov theory (ETDHFB) which includes higher-order effects such as screening 
of the pairing correlation
is proposed. ETDHFB is applied to a fermion system trapped in a harmonic potential 
to test its feasability by comparison with the exact solution. 
With the use of perturbative expressions for the pairing tensor and the two-body density matrix derived from ETDHFB,
the screening effect is investigated for atomic fermion systems and isotopes of tin nuclei.
It is found that the screening effect on the pairing correlation is not significant.
\end{abstract}
\maketitle
\section{Introduction}
The study of higher-order effects on superfluidity has been attracting strong theoretical interests in many fields of physics 
including nuclear physics.
Many-body effects that go beyond the Bardeen-Cooper-Schrieffer theory (BCS) may include the medium polarization known as 
Gorkov and Melik-Barkhudarov (GMB) correction \cite{gorkov}, the self-energy correction, the vertex correction, and so on.
Most calculations for neutron matter \cite{schulze,cao,baldo,gez1} and dilute Fermi gases \cite{gorkov,gez1,heisel,tanizaki} 
show suppression of the pairing correlation due to the medium polarization, whereas studies for finite nuclei 
treating the medium polarization as low-lying vibrations give opposite results \cite{barranco}.
Theoretical studies on the higher-order effects usually start from the generalized gap equation \cite{noz} 
which consists of the particle-particle irreducible kernel and the anomalous propagator, and higher-order corrections are made for these quantities.
The fact that various approaches give contradictory results suggests the necessity of
a consistent microscopic treatment of various higher-order effects on the same footing. 
Monte Carlo calculations \cite{fabrocini1,fabrocini2,gez1,gez2}
and, eventually, exact diagonalisation are certainly consistent approaches 
but restricted to rather small systems (and configuration spaces for the latter) and, thus, have also their limitations. 
It is, therefore, desirable to develop many body techniques which go beyond the standard BCS theory in a systematic way and check their validity for cases where exact solutions can be obtained.

In the present paper we propose an extension of the time-dependent Hartree-Bogoliubov theory (TDHFB) to include higher-order effects.
We formulate the extended TDHFB (ETDHFB) using a truncation scheme similar to that used in the time-dependent density-matrix theory (TDDM) in the normal-fluid regime \cite{WC,GT},
where higher-order reduced density matrices are approximated by lower-order density matrices to truncate the Bogoliubov-Born-Green-Kirkwood-Yvon (BBGKY) hierarchy for
reduced density matrices. TDDM has in the past demonstrated its effectiveness in various applications \cite{GT,pfitz,toh14}
and it can reasonably be assumed that its extension to the superfluid case will show equally good performance.
The advantages of ETDHFB are that it has a direct connection to TDHFB and that various correction terms
are expressed explicitly, contrary to Monte Carlo approaches. 
To show the feasability of ETDHFB,
we apply it to a fermion system trapped in a harmonic potential where comparison with the exact solution 
can be made. Using perturbative expressions for the
pairing tensor and two-body density matrix derived from ETDHFB, we study the screening effect on the
pairing correlation for trapped fermion systems and nuclei of  tin isotopes and make contact with earlier work.\\

The paper is organized as follows. The ETDHFB equations and the perturbative expressions for the pairing tensor and the 
two-body correlation matrix are given in sect. II. 
The obtained results for the trapped fermions and tin isotopes 
are presented in sect. III, and sect. IV is
devoted to the summary.

\section{Formulation}
\subsection{ETDHFB equations}
We consider a Hamiltonian consisting of a one-body part and a two-body interaction :
\begin{eqnarray}
H=\sum_{\alpha\alpha'}\langle\alpha|t|\alpha'\rangle a^\dag_\alpha a_{\alpha'}
+\frac{1}{2}\sum_{\alpha\beta\alpha'\beta'}\langle\alpha\beta|v|\alpha'\beta'\rangle
a^\dag_{\alpha}a^\dag_\beta a_{\beta'}a_{\alpha'},
\nonumber \\
\label{totalh0}
\end{eqnarray}
where $a^\dag_\alpha$ and $a_\alpha$ are the creation and annihilation operators of a fermion in
a time-independent single-particle state $\alpha$.

We first consider the equation of motion for the density matrix ${n}_{\alpha\alpha'}$ which is defined as 
${n}_{\alpha\alpha'}=\langle\Phi(t)|a^\dag_{\alpha'}a_{\alpha}|\Phi(t)\rangle$. Here, $|\Phi(t)\rangle$
is the time-dependent total wavefunction $|\Phi(t)\rangle=\exp(-i(H-\mu \hat{N})/\hbar)|\Phi(0)\rangle$,
where $\hat{N}$ is the number operator and $\mu$ is the chemical potential.
In the equation of motion for the density matrix $i\hbar \dot{n}_{\alpha\alpha'}=\langle\Phi(t)|[a^\dag_{\alpha'}a_{\alpha},H-\mu \hat{N}]|\Phi(t)\rangle$,
there appears a two-body density matrix $\rho_{\alpha\beta\alpha'\beta'}=\langle\Phi(t)|a^\dag_{\alpha'}a^\dag_{\beta'}a_{\beta}a_{\alpha}|\Phi(t)\rangle$.
We decompose it as 
\begin{eqnarray}
\rho_{\alpha\beta\alpha'\beta'}&=&n_{\alpha\alpha'}n_{\beta\beta'}-n_{\alpha\beta'}n_{\beta\alpha'}
\nonumber \\
&+&\kappa_{\alpha\beta}\kappa^*_{\alpha'\beta'}
+{\cal C}_{\alpha\beta\alpha'\beta'}.
\label{rho2}
\end{eqnarray}
Here, $\kappa_{\alpha\beta}$ is the pairing tensor given by $\kappa_{\alpha\beta}=\langle\Phi(t)|a_{\beta}a_{\alpha}|\Phi(t)\rangle$.
The matrix ${\cal C}_{\alpha\beta\alpha'\beta'}$ describes two-body correlations which are not included through the pairing tensor.
In TDHFB the last term in Eq. (\ref{rho2}), that is ${\cal C}_{\alpha\beta\alpha'\beta'}$, is neglected. Similarly, in the equation of motion for the pairing tensor
$i\hbar \dot{\kappa}_{\alpha\beta}=\langle\Phi(t)|[a_{\beta}a_{\alpha},H-\mu \hat{N}]|\Phi(t)\rangle$, there appears
a matrix given by $\langle\Phi(t)|a^\dag_{\alpha'}a_{\gamma}a_{\beta}a_{\alpha}|\Phi(t)\rangle$. We decompose it as 
\begin{eqnarray}
\langle\Phi(t)|a^\dag_{\alpha'}a_{\gamma}a_{\beta}a_{\alpha}|\Phi(t)\rangle&=&n_{\gamma\alpha'}\kappa_{\alpha\beta}
-n_{\beta\alpha'}\kappa_{\alpha\gamma}
\nonumber \\
&+&n_{\alpha\alpha'}\kappa_{\beta\gamma}+K_{\alpha\beta\gamma:\alpha'}.
\label{k2}
\end{eqnarray}
The last term in the above equation is omitted in TDHFB. The matrices 
${\cal C}_{\alpha\beta\alpha'\beta'}$ and $K_{\alpha\beta\gamma:\alpha'}$ describe higher-order effects.
The equation for the density matrix is now extended as 
\begin{eqnarray}
i\hbar \dot{n}_{\alpha\alpha'}&=&
\sum_{\lambda}(\epsilon_{\alpha\lambda}{n}_{\lambda\alpha'}-{n}_{\alpha\lambda}\epsilon_{\lambda\alpha'})
\nonumber \\
&+&\sum_{\lambda}
(\Delta_{\alpha\lambda}\kappa^*_{\alpha'\lambda}-\Delta^*_{\alpha'\lambda}\kappa_{\alpha\lambda}),
\nonumber \\
&+&\sum_{\lambda_1\lambda_2\lambda_3}
[\langle\alpha\lambda_1|v|\lambda_2\lambda_3\rangle {\cal C}_{\lambda_2\lambda_3\alpha'\lambda_1}
\nonumber \\
&-&{\cal C}_{\alpha\lambda_1\lambda_2\lambda_3}\langle\lambda_2\lambda_3|v|\alpha'\lambda_1\rangle],
\label{hfb1}
\end{eqnarray}
where $\epsilon_{\alpha\alpha'}$ is given by
\begin{eqnarray}
\epsilon_{\alpha\alpha'}=\langle\alpha|t|\alpha'\rangle
+\sum_{\lambda_1\lambda_2}
\langle\alpha\lambda_1|v|\alpha'\lambda_2\rangle_A 
n_{\lambda_2\lambda_1},
\label{hf}
\end{eqnarray}
and the pairing potential $\Delta_{\alpha\beta}$ by
\begin{eqnarray}
\Delta_{\alpha\beta}=\frac{1}{2}\sum_{\lambda_1\lambda_2}\langle\alpha\beta|v|\lambda_1\lambda_2\rangle_A\kappa_{\lambda_1\lambda_2}.
\label{gap}
\end{eqnarray}
Here, the subscript $A$ means that the corresponding matrix is antisymmetrized. 
The equation of motion for ${\cal C}_{\alpha\beta\alpha'\beta'}$ is 
given by 
\begin{eqnarray}
i\hbar\dot{\cal C}_{\alpha\beta\alpha'\beta'}&=&
\sum_{\lambda}{(\epsilon_{\alpha\lambda}{\cal C}_{\lambda\beta\alpha'\beta'}
+\epsilon_{\beta\lambda}{\cal C}_{\alpha\lambda\alpha'\beta'}}
\nonumber \\
&-&{\epsilon_{\lambda\alpha'}{\cal C}_{\alpha\beta\lambda\beta'}
-\epsilon_{\lambda\beta'}{\cal C}_{\alpha\beta\alpha'\lambda})}
\nonumber \\ 
&+&{B_{\alpha\beta\alpha'\beta'}}+P_{\alpha\beta\alpha'\beta'}+H_{\alpha\beta\alpha'\beta'}
\nonumber \\ 
&+&S_{\alpha\beta\alpha'\beta'}+T_{\alpha\beta\alpha'\beta'}.
\label{hfbc}
\end{eqnarray}
In order to close the coupled chain of equations of motion, we approximated the matrix $\langle\Phi(t)|a^\dag_{\alpha'}a^\dag_{\beta'}a^\dag_{\gamma'}a_{\gamma}a_{\beta}a_{\alpha}|\Phi(t)\rangle$
by antisymmetrized product combinations of 
$n_{\alpha\alpha'}$, $\kappa_{\alpha\beta}$, ${\cal C}_{\alpha\beta\alpha'\beta'}$ and $K_{\alpha\beta\gamma:\alpha'}$
such as $n_{\alpha\alpha'}n_{\beta\beta'}n_{\gamma\gamma'}$, $n_{\alpha\alpha'}C_{\beta\gamma\beta'\gamma'}$, $n_{\alpha\alpha'}\kappa_{\beta\gamma}\kappa^*_{\beta'\gamma'}$,
$\kappa_{\alpha\beta}K^*_{\alpha'\beta'\gamma':\gamma}$ and $\kappa^*_{\alpha'\beta'}K_{\alpha\beta\gamma:\gamma'}$.
In Eq. (\ref{hfbc}) $B_{\alpha\beta\alpha'\beta'}$ describes the two particle (2p) - two hole (2h) and 2h-2p excitations,
$P_{\alpha\beta\alpha'\beta'}$ 
p-p (and h-h) correlations which are not included in the pairing tensor, and 
$H_{\alpha\beta\alpha'\beta'}$ p-h correlations. 
The terms in $S_{\alpha\beta\alpha'\beta'}$ and 
$T_{\alpha\beta\alpha'\beta'}$ express the coupling to $\kappa_{\alpha\beta}$ and $K_{\alpha\beta\gamma:\alpha'}$, respectively.
The expressions for the matrices in Eq. (\ref{hfbc}) are given in Appendix A.
The equation for ${\cal C}_{\alpha\beta\alpha'\beta'}$ without $S_{\alpha\beta\alpha'\beta'}$ and 
$T_{\alpha\beta\alpha'\beta'}$ are the same as that in TDDM \cite{GT}.
Since the total wavefunction $|\Phi(t)\rangle$ is not an eigenstate of the number operator, the couplings to $\kappa_{\alpha\beta}$ and $K_{\alpha\beta\gamma:\alpha'}$
appear in Eq. (\ref{hfbc}).

The equation for the pairing tensor is also extended so that
\begin{eqnarray}
i\hbar\dot{\kappa}_{\alpha\beta}&=&\sum_{\lambda}(\tilde{\epsilon}_{\alpha\lambda}{\kappa}_{\lambda\beta}+\tilde{\epsilon}_{\beta\lambda}{\kappa}_{\alpha\lambda})
+\Delta_{\alpha\beta}
\nonumber \\
&+&\sum_{\lambda}
(\Delta_{\beta\lambda}n_{\alpha\lambda}-\Delta_{\alpha\lambda}n_{\beta\lambda})
\nonumber \\
&-&\sum_{\lambda_1\lambda_2\lambda_3}(\langle\alpha\lambda_1|v|\lambda_2\lambda_3\rangle K_{\beta\lambda_2\lambda_3:\lambda_1}
\nonumber \\
&+&\langle\lambda_1\beta|v|\lambda_2\lambda_3\rangle K_{\alpha\lambda_2\lambda_3:\lambda_1}),
\label{hfb2}
\end{eqnarray}
where 
$\tilde{\epsilon}_{\alpha\alpha'}=\epsilon_{\alpha\alpha'}-\mu\delta_{\alpha\alpha'}$.
The equation for $K_{\alpha\beta\gamma:\alpha'}$ is written as
\begin{eqnarray}
i\hbar\dot{K}_{\alpha\beta\gamma:\alpha'}&=&\sum_{\lambda}{(\tilde{\epsilon}_{\alpha\lambda}{K}_{\lambda\beta\gamma:\alpha'}+\tilde{\epsilon}_{\beta\lambda}{K}_{\alpha\lambda\gamma:\alpha'}}
\nonumber \\
&+&{\tilde{\epsilon}_{\gamma\lambda}{K}_{\alpha\beta\lambda:\alpha'}-\tilde{\epsilon}_{\lambda\alpha'}{K}_{\alpha\beta\gamma:\lambda})}
\nonumber \\
&+&{D_{\alpha\beta\gamma:\alpha'}}
+ E_{\alpha\beta\gamma:\alpha'},
\nonumber \\
&+& F_{\alpha\beta\gamma:\alpha'}
+ G_{\alpha\beta\gamma:\alpha'}.
\label{hfb3}
\end{eqnarray}
We approximated the matrix
$\langle\Phi(t)|a^\dag_{\alpha'}a^\dag_{\beta'}a_{\delta}a_{\gamma}a_{\beta}a_{\alpha}|\Phi(t)\rangle$ by antisymmetrized product combinations of 
$n_{\alpha\alpha'}$, $\kappa_{\alpha\beta}$, ${\cal C}_{\alpha\beta\alpha'\beta'}$ and $K_{\alpha\beta\gamma:\alpha'}$.
The terms in
$D_{\alpha\beta\gamma:\alpha'}$ and $E_{\alpha\beta\gamma:\alpha'}$ describe the coupling to the pairing tensor 
and to the product of three pairing tensors, respectively.
The terms in $F_{\alpha\beta\gamma:\alpha'}$ describe correlations involving  $K_{\alpha\beta\gamma:\alpha'}$.
The coupling to ${\cal C}_{\alpha\beta\alpha'\beta'}$ is contained in $G_{\alpha\beta\gamma:\alpha'}$. 
The matrices in Eq. (\ref{hfb3}) are given in Appendix A.
Equations (\ref{hfb1}) and (\ref{hfb2}) may be written in matrix form as in TDHFB
\begin{eqnarray}
i\hbar\dot{\cal R}-[{\cal H},{\cal R}]=[{\cal V},{\cal K}],
\end{eqnarray}
where in obvious notation
\begin{eqnarray}
{\cal R}&=&\left(
\begin{array}{cc}
n& \kappa
\\
-\kappa^*& 1-n^*
\end{array}
\right),\\
{\cal H}&=&\left(
\begin{array}{cc}
\epsilon& \Delta\\
-\Delta^*& -\epsilon^*
\end{array}
\right),
\\
{\cal K}&=&\left(
\begin{array}{cc}
{\cal C}& K\\
-K^*& -{\cal C}^*
\end{array}
\right), 
\\
{\cal V}&=&\left(
\begin{array}{cc}
v& 0\\
0& -v^*
\end{array}
\right).
\end{eqnarray}

The ETDHFB equation Eq. (\ref{hfb1})
conserves on average the total number of particles $N=\sum_\alpha n_{\alpha\alpha}$ as is easily shown by taking the trace of Eq. (\ref{hfb1}).
The total energy $E_{\rm tot}$ 
\begin{eqnarray}
E_{\rm tot}&=&\sum_{\alpha}\epsilon_\alpha n_{\alpha\alpha}
\nonumber \\
&+&\frac{1}{2}\sum_{\alpha\beta\alpha'\beta'}\langle\alpha\beta|v|\alpha'\beta'\rangle
\rho_{\alpha'\beta'\alpha\beta}
\label{etot}
\end{eqnarray}
may be divided into the mean-field energy $E_{\rm MF}$, the pairing energy $E_{\rm pair}$ 
and the correlation energy $E_{\rm corr}$ given by 
\begin{eqnarray}
E_{\rm MF}&=&\sum_{\alpha}\epsilon_\alpha n_{\alpha\alpha}
\nonumber \\
&+&\frac{1}{2}\sum_{\alpha\beta\alpha'\beta'}\langle\alpha\beta|v|\alpha'\beta'\rangle_A
n_{\alpha'\alpha}n_{\beta'\beta},
\label{mean}
\end{eqnarray}
\begin{eqnarray}
E_{\rm pair}=\frac{1}{2}\sum_{\alpha\beta}\Delta_{\alpha\beta}\kappa^*_{\alpha\beta},
\label{epair}
\end{eqnarray}
\begin{eqnarray}
E_{\rm corr}=\frac{1}{2}\sum_{\alpha\beta\alpha'\beta'}\langle\alpha\beta|v|\alpha'\beta'\rangle
{\cal C}_{\alpha'\beta'\alpha\beta}.
\label{ecorr}
\end{eqnarray}
To conserve $E_{\rm tot}$, we need all
ETDHFB equations Eqs. (\ref{hfb1}), (\ref{hfbc}), (\ref{hfb2}), and (\ref{hfb3}).

\subsection{Perturbative expression}
To understand various higher-order effects included in ETDHFB, we derive perturbative expressions
for the pairing tensor and the two-body correlation matrix and show how the screening effect is treated in ETDHFB. 
\subsubsection{Pairing tensor}
\begin{figure}
\begin{center}
\includegraphics[height=6cm]{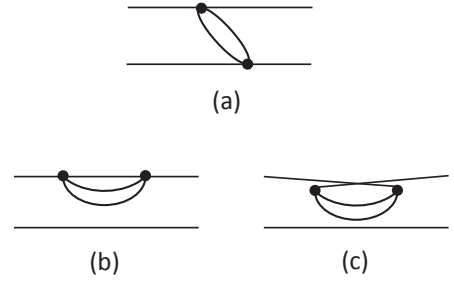}
\end{center}
\caption{Schematic presentation of higher-order effects: (a) Screening effect, and (b) and (c)
self-energy correction. Lines depict single-particle states and dots the residual interaction.}
\label{etdhfbfig}
\end{figure}
First we derive a perturbative expression for the pairing tensor using the equations of motion of ETDHFB.
Since the $F$ and $G$ terms in Eq. (\ref{hfb3}) which include ${K}_{\alpha\beta\gamma:\alpha'}$
and ${\cal C}_{\alpha\beta\alpha'\beta'}$ are of higher order, we consider only the $D$ term and
assume
that the single-particle energy $\tilde{\epsilon}_{\alpha\alpha'}$,
the density matrix $n_{\alpha\alpha'}$ and the pairing tensor $\kappa_{\alpha\beta}$ are diagonal:
$\tilde{\epsilon}_{\alpha\alpha'}=\tilde{\epsilon}_\alpha\delta_{\alpha\alpha'}$,
$n_{\alpha\alpha'}=n_\alpha\delta_{\alpha\alpha'}$ and $\kappa_{\alpha\beta}=\kappa_{\alpha}\delta_{\beta\bar{\alpha}}$
where $\bar{\alpha}$ stands for the time-reversal state of $\alpha$. 
The $E$ term is also neglected because $\kappa_\alpha\kappa_\beta$ is small for the p-h transition where $\bar{n}_\alpha n_\beta\approx 1$.
%\textcolor{blue}{ Is this really true? for $\alpha = \beta$, we have $u^2v^2$ for both expressions, right?}
Then
Eq. (\ref{hfb3}) is written as 
\begin{eqnarray}
i\hbar\dot{K}_{\alpha\beta\gamma:\alpha'}&\approx&(\tilde{\epsilon}_{\alpha}+\tilde{\epsilon}_{\beta}
+\tilde{\epsilon}_{\gamma}-\tilde{\epsilon}_{\alpha'}){K}_{\alpha\beta\gamma:\alpha'}
\nonumber \\
&-&\langle\alpha\beta|v|\alpha'\bar{\gamma}\rangle_A(\bar{n}_\alpha\bar{n}_\beta n_{\alpha'}+n_\alpha n_\beta\bar{n}_{\alpha'})\kappa_\gamma
\nonumber \\
&-&\langle\beta\gamma|v|\alpha'\bar{\alpha}\rangle_A(\bar{n}_\beta\bar{n}_\gamma n_{\alpha'}+n_\beta n_\gamma\bar{n}_{\alpha'})\kappa_\alpha
\nonumber \\
&+&\langle\alpha\gamma|v|\alpha'\bar{\beta}\rangle_A(\bar{n}_\alpha\bar{n}_\gamma n_{\alpha'}+n_\alpha n_\gamma\bar{n}_{\alpha'})\kappa_\beta,
\nonumber \\
\label{K}
\end{eqnarray}
where $\bar{n}_\alpha=1-n_\alpha$.
The stationary condition $\dot{K}_{\alpha\beta\gamma:\alpha'}=0$ gives a perturbative expression for ${K}_{\alpha\beta\gamma:\alpha'}$.
Inserting it into Eq. (\ref{hfb2}) and using the stationary condition $\dot{\kappa}_{\alpha\beta}=0$, 
we can write the equation for the pairing tensor as
\begin{eqnarray}
2\tilde{\epsilon}_\alpha\kappa_\alpha&\approx&
-(1-2n_{\alpha})\Delta_{\alpha}
\nonumber \\
&+&2\sum_{\lambda_1\lambda_2\lambda_3}\langle\alpha\lambda_1|v|\lambda_2\lambda_3\rangle_A
\nonumber \\
&\times&\frac{\bar{n}_{\alpha}\bar{n}_{\lambda_3}{n}_{\lambda_1}+{n}_{\alpha}{n}_{\lambda_3}\bar{n}_{\lambda_1}}
{\tilde{\epsilon}_\alpha+\tilde{\epsilon}_{\lambda_2}+\tilde{\epsilon}_{\lambda_3}-\tilde{\epsilon}_{\lambda_1}}
\nonumber \\
&\times&\langle\bar{\alpha}\lambda_3|v|\bar{\lambda}_2\lambda_1\rangle_A
\kappa_{\lambda_2}
\nonumber \\
&+&
\sum_{\lambda_1\lambda_2\lambda_3}\langle\alpha\lambda_1|v|\lambda_2\lambda_3\rangle_A
\nonumber \\
&\times&\frac{\bar{n}_{\lambda_2}\bar{n}_{\lambda_3}{n}_{\lambda_1}+{n}_{\lambda_2}{n}_{\lambda_3}\bar{n}_{\lambda_1}}
{\tilde{\epsilon}_\alpha+\tilde{\epsilon}_{\lambda_2}+\tilde{\epsilon}_{\lambda_3}-\tilde{\epsilon}_{\lambda_1}}
\nonumber \\
&\times&
\langle\lambda_2\lambda_3|v|\alpha\lambda_1\rangle_A
\kappa_\alpha.
\nonumber \\
\label{pertk0}
\end{eqnarray}
The second term on the right-hand side can be interpreted as a correction to $\Delta_{\alpha}$
because it contains the sum over the pairing tensor as the pair potential does.
The corresponding diagram is shown in Fig. \ref{etdhfbfig}(a). 
We call it the screening term because a similar process has been shown
responsible for the screening of the pairing correlation \cite{gorkov,schulze,cao,heisel}. 
The last term on the right-hand side of Eq. (\ref{pertk0}) can be interpreted as the self-energy correction to the single-particle energies 
$2\tilde{\epsilon}_\alpha$ because it is proportional to $\kappa_\alpha$ as the term on the left-hand side of Eq. (\ref{pertk0}).
The corresponding diagrams are schematically shown in Fig. \ref{etdhfbfig} ((b) and (c)). 
Using the BCS relations
\begin{eqnarray}
n_\alpha&=&v^2_\alpha=\frac{1}{2}(1-\frac{\tilde{\epsilon}_\alpha}{E_\alpha}),
\\
\kappa^0_\alpha&=&v_\alpha u_\alpha=-\frac{1}{2\tilde{\epsilon}_\alpha}
(1-2n_\alpha)\Delta_\alpha
%\nonumber \\
=-\frac{1}{2E_\alpha}\Delta_\alpha,
\label{bcs}
\end{eqnarray}
where $\kappa^0_\alpha$ is the pairing tensor in BCS and
$E_\alpha$ is the quasi-particle energy $E_\alpha=\sqrt{\tilde{\epsilon}^2_\alpha+\Delta^2_\alpha}$,
and expressing $2\tilde{\epsilon}_\alpha\kappa_\alpha-2n_\alpha\Delta_\alpha$ 
%\textcolor{blue}{ must it not be a $+$ sign?} 
as $2E_\alpha\kappa_\alpha$, we
sovle Eq. (\ref{pertk0}) for the pairing tensor
\begin{eqnarray}
\kappa_\alpha&\approx&\kappa^0_\alpha+
\frac{1}{E_\alpha}
\sum_{\lambda_1\lambda_2\lambda_3}\langle\alpha\lambda_1|v|\lambda_2\lambda_3\rangle_A
\nonumber \\
&\times&\frac{\bar{n}_{\alpha}\bar{n}_{\lambda_3}{n}_{\lambda_1}+{n}_{\alpha}{n}_{\lambda_3}\bar{n}_{\lambda_1}}
{\tilde{\epsilon}_\alpha+\tilde{\epsilon}_{\lambda_2}+\tilde{\epsilon}_{\lambda_3}-\tilde{\epsilon}_{\lambda_1}}
\nonumber \\
&\times&\langle\bar{\alpha}\lambda_3|v|\bar{\lambda}_2\lambda_1\rangle_A
\kappa^0_{\lambda_2}
\nonumber \\
&+&\frac{\kappa^0_\alpha}{2E_\alpha}\sum_{\lambda_1\lambda_2\lambda_3}\langle\alpha\lambda_1|v|\lambda_2\lambda_3\rangle_A
\nonumber \\
&\times&\frac{\bar{n}_{\lambda_2}\bar{n}_{\lambda_3}{n}_{\lambda_1}+{n}_{\lambda_2}{n}_{\lambda_3}\bar{n}_{\lambda_1}}
{\tilde{\epsilon}_\alpha+\tilde{\epsilon}_{\lambda_2}+\tilde{\epsilon}_{\lambda_3}-\tilde{\epsilon}_{\lambda_1}}
\langle\lambda_2\lambda_3|v|\alpha\lambda_1\rangle_A.
\label{pertk}
\end{eqnarray}
Inserting the above expression for $\kappa_\alpha$
into Eq. (\ref{gap}), we obtain the pair potential and also the correction to the pairing energy Eq. (\ref{epair}).
The spin state of the single-particle state $\lambda_2$ in the screening term of Eq. (\ref{pertk}) must be the same as $\bar{\alpha}$.
Therefore, the screening effect is compensated by
the self-energy correction.
The effects of the mean-field
contribution and the partial
occupation of the single-particle states are also included through $\tilde{\epsilon}_\alpha$ and the Pauli blocking factor, respectively.

\subsubsection{Relation to other perturbative approaches}
Next we discuss the relation of our perturbative formulation and  
the expression used in 
Refs. \cite{schulze,cao,heisel} to study the screening effect. 
The latter is related to the self-energy $\Sigma_{1\alpha}$
of the Gorkov Green's function (see Appendix B), where
\begin{eqnarray}
\Sigma_{1\alpha}&=&\sum_{\lambda_1\lambda_2\lambda_3}
\langle\alpha\lambda_1|v|\lambda_2\lambda_3\rangle_A
\nonumber \\
&\times&\frac{n_{\lambda_3}-n_{\lambda_1}}{\omega_\mu+\tilde{\epsilon}_{\lambda_2}+\tilde{\epsilon}_{\lambda_3}-\tilde{\epsilon}_{\lambda_1}}
\langle\bar{\alpha}\lambda_3|v|\bar{\lambda}_2\lambda_1\rangle_A\kappa_{\lambda_2}
\label{x11}.
\end{eqnarray}
We focus on the second term on the right-hand side of Eq. (\ref{pertk0}) and neglect for the purpose of discussion for the moment the last term (the self-energy correction).
Rewriting 
the numerator of the second term as 
\begin{eqnarray}
\bar{n}_{\alpha}\bar{n}_{\lambda_3}{n}_{\lambda_1}&+&{n}_{\alpha}{n}_{\lambda_3}\bar{n}_{\lambda_1}=
\nonumber \\
\bar{n}_{\lambda_3}{n}_{\lambda_1}&+&{n}_{\alpha}({n}_{\lambda_3}\bar{n}_{\lambda_1}-\bar{n}_{\lambda_3}{n}_{\lambda_1})
\nonumber \\
=\bar{n}_{\lambda_3}{n}_{\lambda_1}&+&{n}_{\alpha}({n}_{\lambda_3}-{n}_{\lambda_1}),
\end{eqnarray}
we can express
Eq. (\ref{pertk0}) without the self-energy contribution such that
\begin{eqnarray}
2\tilde{\epsilon}_\alpha\kappa_\alpha
=-(\Delta_{\alpha}+\Sigma_\alpha)+2n_{\alpha}{\Delta'}_{\alpha},
\label{exgap2}
\end{eqnarray}
where
\begin{eqnarray}
\Sigma_\alpha&=&-2\sum_{\lambda_1\lambda_2\lambda_3}\langle\alpha\lambda_1|v|\lambda_2\lambda_3\rangle_A
\nonumber \\
&\times&\frac{\bar{n}_{\lambda_3}{n}_{\lambda_1}}
{\tilde{\epsilon}_\alpha+\tilde{\epsilon}_{\lambda_2}+\tilde{\epsilon}_{\lambda_3}-\tilde{\epsilon}_{\lambda_1}}
\nonumber \\
&\times&\langle\bar{\alpha}\lambda_3|v|\bar{\lambda}_2\lambda_1\rangle_A
\kappa_{\lambda_2}
\end{eqnarray}
and ${\Delta'}_{\alpha}=\Delta_\alpha+\Sigma_{1\alpha}(\omega_\mu=\tilde{\epsilon}_\alpha)$.
If we consider the single-particle state near $\mu$ ($\tilde{\epsilon}_\alpha\approx 0$) and assume that the pairing tensor for the single-particle state
around $\mu$ dominates (this means also $\tilde{\epsilon}_{\lambda_2}\approx 0$), $\Sigma_{1\alpha}$ is simplified to
\begin{eqnarray}
\Sigma_{1\alpha}(\omega_\mu\approx0)&\approx&\sum_{\lambda_1\lambda_2\lambda_3}
\langle\alpha\lambda_1|v|\lambda_2\lambda_3\rangle_A
\frac{n_{\lambda_3}-n_{\lambda_1}}{\tilde{\epsilon}_{\lambda_3}-\tilde{\epsilon}_{\lambda_1}}
\nonumber \\
&\times&\langle\bar{\alpha}\lambda_3|v|\bar{\lambda_2}\lambda_1\rangle_A\kappa_{\lambda_2}
\label{F1}
\end{eqnarray}
and $\Sigma_\alpha$ is also given by
\begin{eqnarray}
\Sigma_\alpha&\approx&-2\sum_{\lambda_1\lambda_2\lambda_3}\langle\alpha\lambda_1|v|\lambda_2\lambda_3\rangle_A
\nonumber \\
&\times&\frac{\bar{n}_{\lambda_3}{n}_{\lambda_1}}
{\tilde{\epsilon}_{\lambda_3}-\tilde{\epsilon}_{\lambda_1}}
\nonumber \\
&\times&\langle\bar{\alpha}\lambda_3|v|\bar{\lambda}_2\lambda_1\rangle_A
\kappa_{\lambda_2}
\nonumber \\
&=&\sum_{\lambda_1\lambda_2\lambda_3}\langle\alpha\lambda_1|v|\lambda_2\lambda_3\rangle_A
\nonumber \\
&\times&\frac{{n}_{\lambda_3}-{n}_{\lambda_1}}
{\tilde{\epsilon}_{\lambda_3}-\tilde{\epsilon}_{\lambda_1}}
\nonumber \\
&\times&\langle\bar{\alpha}\lambda_3|v|\bar{\lambda}_2\lambda_1\rangle_A
\kappa_{\lambda_2}
\end{eqnarray}
In this limit the relation
$\Sigma_\alpha\approx\Sigma_{1\alpha}$ holds and 
Eq. (\ref{exgap2}) is written as
\begin{eqnarray}
2\tilde{\epsilon}_\alpha\kappa_\alpha
=-(1-2n_{\alpha})\Delta'_{\alpha}.
\label{exgap0}
\end{eqnarray}
If Eq. (\ref{exgap0}) is treated as the BCS equation for $\kappa_\alpha$, we obtain the modified quasi-particle energy 
$E'_\alpha=\sqrt{\tilde{\epsilon}_\alpha^2+{\Delta'}_\alpha^2}$
and pairing tensor $\kappa'_\alpha=-{\Delta'}_{\alpha}/2E'_\alpha$.
The modified gap equation is written as 
\begin{eqnarray}
{\Delta'}_\alpha
=-\sum_\lambda F_{\alpha:\lambda} \frac{{\Delta'}_{\lambda}}{2E'_\lambda},
\label{exgap1}
\end{eqnarray}
where $F_{\alpha:\lambda}$ is given by
\begin{eqnarray}
F_{\alpha:\lambda}&=&\frac{1}{2}\langle\alpha\bar{\alpha}|v|\lambda\bar{\lambda}\rangle_A
\nonumber \\
&+&\sum_{\lambda_1\lambda_2}\langle\alpha\lambda_1|v|\lambda\lambda_2\rangle_A
\nonumber \\
&\times&\frac{{n}_{\lambda_2}-{n}_{\lambda_1}}
{\tilde{\epsilon}_{\lambda_2}-\tilde{\epsilon}_{\lambda_1}}
\nonumber \\
&\times&\langle\bar{\alpha}\lambda_2|v|\bar{\lambda}\lambda_1\rangle_A.
\label{F}
\end{eqnarray}
When we further assume that $n_\alpha=0$ or 1, we arrive at the perturbative expression of Refs. \cite{schulze,cao,heisel}. 
For a simple contact interaction $g\delta^3({\bm r}-{\bm r'})$
Eq. (\ref{F1}) always gives a positive value (screening).
The difference between Eqs. (\ref{exgap2}) and (\ref{exgap0}) stems from the difference in the occupation factors in the numerator between Eqs. (\ref{pertk0}) and
(\ref{x11}). The occupation factor in Eq. (\ref{pertk0}) describes 
a blocking effect of the ph excitation caused by the existence of another particle. As discussed, this difference may be small if pairing is concentrated to states close to the Fermi level (weak coupling).

\subsubsection{Two-body correlation matrix}
Now we consider the corrections to the correlation energy Eq. (\ref{ecorr}) which are given by
the pertubative expression for the two-body correlation matrix. In Eq. (\ref{hfbc})
the terms in $P_{\alpha\beta\alpha'\beta'}$ and $H_{\alpha\beta\alpha'\beta'}$ contain ${\cal C}_{\alpha\beta\alpha'\beta'}$, and
$T_{\alpha\beta\alpha'\beta'}$ includes $K_{\alpha\beta\gamma:\alpha'}$. Therefore, the lowest-order corrections are from
$B_{\alpha\beta\alpha'\beta'}$ and $S_{\alpha\beta\alpha'\beta'}$.
The pertubative expression for ${\cal C}_{\alpha\beta\alpha'\beta'}$ obtained using only the terms in $S_{\alpha\beta\alpha'\beta'}$
in Eq. (\ref{hfbc}) is given by 
\begin{eqnarray}
{\cal C}_{1\alpha\beta\alpha'\beta'}&=&\frac{1}
{\epsilon_\alpha+\epsilon_\beta-\epsilon_{\alpha'}-\epsilon_{\beta'}}
\nonumber \\
&\times&[
\langle\alpha\bar{\beta'}|v|\alpha'\bar{\beta}\rangle_A\kappa^0_\beta\kappa^{0*}_{\beta'}(n_\alpha-n_{\alpha'})
\nonumber \\
&+&
\langle\beta\bar{\alpha'}|v|\beta'\bar{\alpha}\rangle_A\kappa^0_\alpha\kappa^{0*}_{\alpha'}(n_\beta-n_{\beta'})
\nonumber \\
&-&
\langle\alpha\bar{\alpha'}|v|\beta'\bar{\beta}\rangle_A\kappa^0_\beta\kappa^{0*}_{\alpha'}(n_\alpha-n_{\beta'})
\nonumber \\
&-&
\langle\beta\bar{\beta'}|v|\alpha'\bar{\alpha}\rangle_A\kappa^0_\alpha\kappa^{0*}_{\beta'}(n_\beta-n_{\alpha'})].
\label{pertc1}
\end{eqnarray} 
The perturbative expression for ${\cal C}_{\alpha\beta\alpha'\beta'}$ obtained from Eq. (\ref{hfbc}) with only the $B_{\alpha\beta\alpha'\beta'}$ 
is written as
\begin{eqnarray}
{\cal C}_{2\alpha\beta\alpha'\beta'}&=&-\frac{\langle\alpha\beta|v|\alpha'\beta'\rangle_A}
{\epsilon_\alpha+\epsilon_\beta-\epsilon_{\alpha'}-\epsilon_{\beta'}}
\nonumber \\
&\times&(\bar{n}_\alpha\bar{n}_\beta n_{\alpha'}n_{\beta'}-{n}_\alpha{n}_\beta \bar{n}_{\alpha'}\bar{n}_{\beta'}),
\label{pertc}
\end{eqnarray}
which describes the 2p-2h and 2h-2p excitations.
The corrections to the correlation energy obtained from ${\cal C}_{1\alpha\beta\alpha'\beta'}$ and ${\cal C}_{2\alpha\beta\alpha'\beta'}$
are related to the self-energies 
$\Sigma_{1\alpha}$ (Eq. (\ref{x11})) and $\Sigma_{2\alpha}$ of the Gorkov Green's function (see Appendix B), where
\begin{eqnarray}
\Sigma_{2\alpha}&=&-\frac{1}{2}\sum_{\lambda\lambda_1\lambda_2\lambda_3}
\langle\alpha\lambda_1|v|\lambda_2\lambda_3\rangle_A
\frac{\bar{n}_{\lambda_2}\bar{n}_{\lambda_3}n_{\lambda_1}+n_{\lambda_2}n_{\lambda_3}\bar{n}_{\lambda_1}}
{\omega_\mu+\tilde{\epsilon}_{\lambda_2}+\tilde{\epsilon}_{\lambda_3}-\tilde{\epsilon}_{\lambda_1}}
\nonumber \\
&\times&\langle\lambda_2\lambda_3|v|\alpha\lambda_1\rangle_A.
\label{x12}
\end{eqnarray}
The self-energy $\Sigma_{1\alpha}$ describes a correction to the pair potential $\Delta_\alpha$, similarly
to the screening term in Eq. (\ref{pertk}), whereas $\Sigma_{2\alpha}$ is a correction to the
mean-field potential as is the case of the normal single-particle Green's function.
The correlation energy obtained from ${\cal C}_{1\alpha\beta\alpha'\beta'}$ corresponds to the contribution of 
$\Sigma_{1\alpha}$ to the total energy
because it is written as $\sum_\alpha\Sigma_{1\alpha}\kappa^*_\alpha$, whereas
the correlation energy obtained from 
${\cal C}_{2\alpha\beta\alpha'\beta'}$ corresponds to
the contribution of $\Sigma_{2\alpha}$. 
The correlation energy obtained from ${\cal C}_{2\alpha\beta\alpha'\beta'}$ gives a significant correction to the BCS total energy
in the case of the pairing Hamiltonian \cite{rich,sandu,lacroix}.

%In our perturbative formulation the screening effect expressed by Eq. (\ref{x13}) is included
%through the correlation matrix Eq. (\ref{pertc1}) which contains the pairing tensor.

\section{Numerical results}

\subsection{Trapped fermions}
First we consider a system of fermions with spin one half, 
which is trapped in a spherically symmetric harmonic potential with
frequency $\omega$. The system is described by the Hamiltonian 
\begin{eqnarray}
H=\sum_\alpha\epsilon_\alpha a^\dag_\alpha a_\alpha
+\frac{1}{2}\sum_{\alpha\beta\alpha'\beta'}\langle\alpha\beta|v|\alpha'\beta'\rangle
a^\dag_{\alpha}a^\dag_\beta a_{\beta'}a_{\alpha'},
\nonumber \\
\label{totalH}
\end{eqnarray}
where $a^\dag_\alpha$ and $a_\alpha$ are the creation and annihilation operators of an atom at
a harmonic oscillator state $\alpha$ 
corresponding to the trapping potential $V(r)=m\omega^2r^2/2$ and
$\epsilon_\alpha=\hbar\omega(n+3/2)$ with $n=0,~1,~2,....$.
We assume that $\alpha$ contains the spin quantum number $\sigma$.
In Eq. (\ref{totalH}) $\langle\alpha\beta|v|\alpha'\beta'\rangle$ is the matrix element of
an attractive contact interaction
$v({\bm r}-{\bm r'})=g\delta^3({\bm r}-{\bm r'})$. 

We consider a system consisting of six fermions whose non-interacting configuration consists of the partially filled $1p$ state.
Besides a trap with a small number of cold atoms, our system may correspond to neutrons in carbon isotopes.
For numerical reasons we only can handle a very restricted spaces and small number of particles, since we want to compare with exact solutions.
Using a limited number of the single-particle states, the $1s$, $1p$, $1d$ and $2s$ states, we obtain the ground states
in the Hartree-Fock-Bogoliubov (HFB) theory and the ETDHFB theory (Eqs. (\ref{hfb1}), (\ref{hfbc}), (\ref{hfb2}), and (\ref{hfb3}) together with the expressions given in Appendix A),
and compare with the exact solution obtained from the diagonalization of the Hamiltonian using the same single-particle space.
The ground state in ETDHFB is obtained using an
adiabatic method \cite{adiabatic}: Starting from the HFB ground state, we solve the coupled set of the ETDHFB equations 
by gradually increasing the
residual interaction $g'=g\times t/T$.
This method is motivated by the Gell-Mann-Low theorem \cite{gell} and has often been used to obtain approximate
ground states \cite{pfitz}. To suppress oscillating components which come from the mixing
of excited states, we must take large $T$: We use $T=4\times2\pi/\omega$.
It has been pointed out \cite{ts14} that TDDM with all components of ${\cal C}_{\alpha\beta\alpha'\beta'}$
overestimates two-body correlations and theoretical arguments have been given that the exclusion of the ph-ph components ${\cal C}_{php'h'}$ is more consistent leading to
good agreement with the exact solutions of solvable models.
Therefore, we discard the ph-ph components between the $1s$ state and the $2s$ and $1d$ states.

\begin{figure}
\begin{center}
\includegraphics[height=6cm]{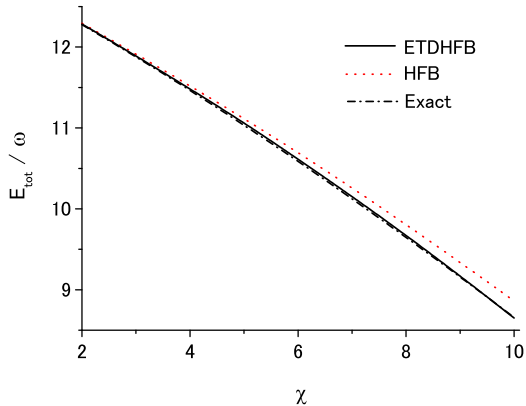}
\end{center}
\caption{(Color online) Total energy as a function of $\chi$ calculated in ETDHFB (solid line). 
The dotted line depicts the results in HFB. The exact solutions are given by the dot-dashed line. }
\label{ehfbtot}
\end{figure}
\begin{figure}
\begin{center}
\includegraphics[height=6cm]{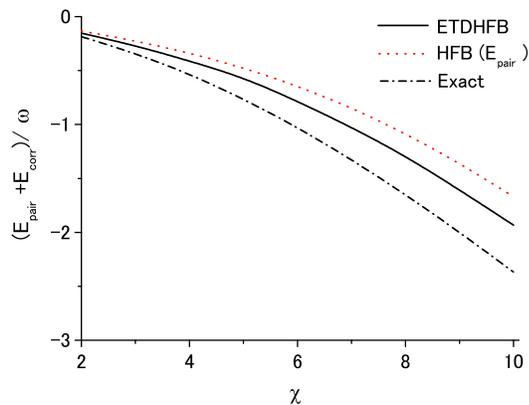}
\end{center}
\caption{(Color online) Sum $E_{\rm pair}+E_{\rm corr}$ as as a function of $\chi$ calculated in ETDHFB (solid line). 
The dot-dashed line depicts the exact solutions. $E_{\rm pair}$ in HFB is shown with the dotted line.}
\label{ehfbcorr}
\end{figure}

\begin{figure}
\begin{center}
\includegraphics[height=6cm]{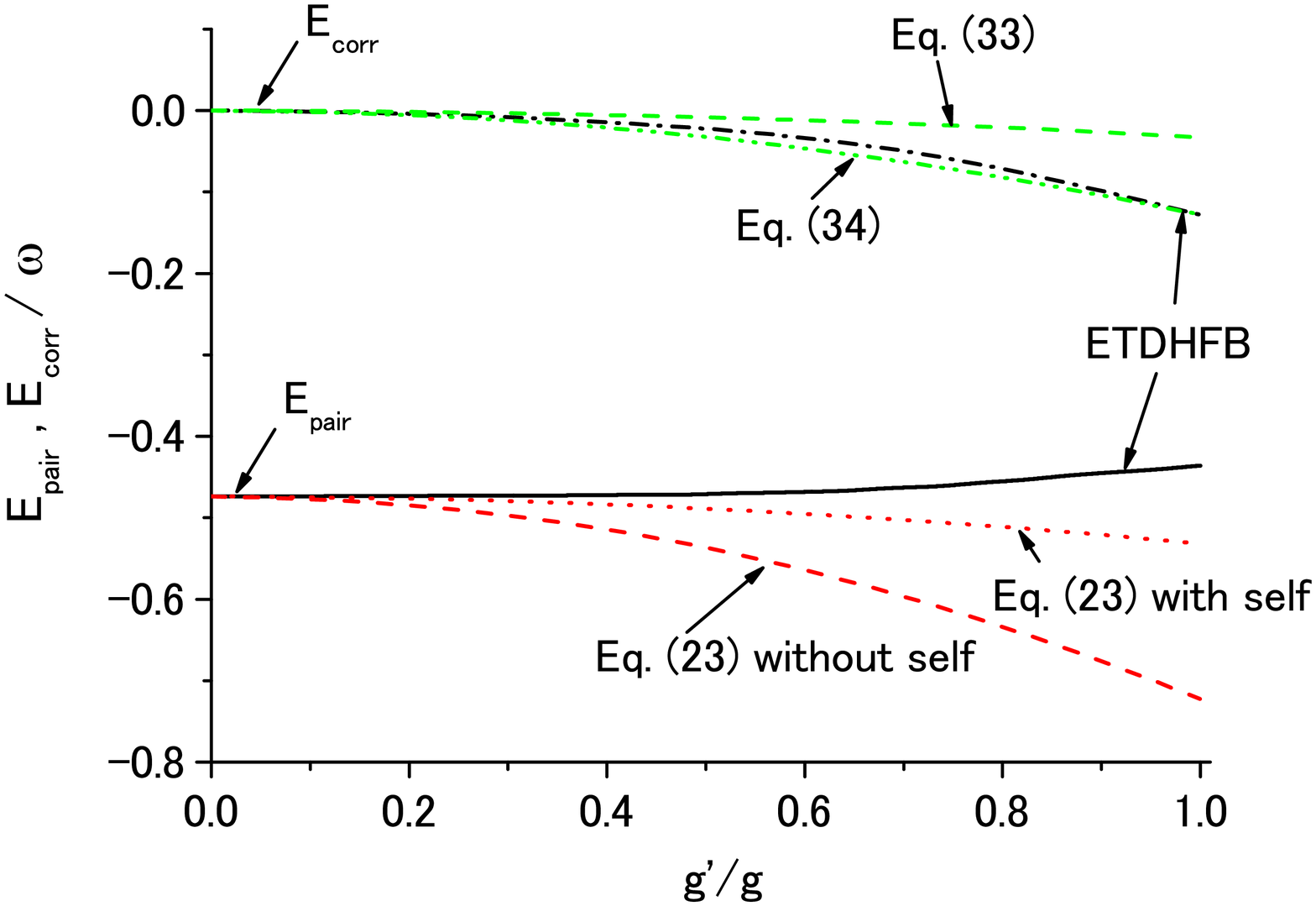}
\end{center}
\caption{(Color online) Pairing energy $E_{\rm pair}$ (solid line) and $E_{\rm corr}$ (dot-dashed line) 
calculated in ETDHFB for $\chi=5$ 
as a function of $g'/g$. 
The correlation energy calculated using the perturbative expression for the two-body correlation matrix
Eq. (\ref{pertc}) is shown with the green (gray) dot-dashed line.
The correlation energy obtained from the two-body correlation matrix Eq. (\ref{pertc1}) is also shown
with the green (gray) dashed line.
The dotted and dashed lines depict the results of
the perturbative approach Eq. (\ref{pertk}) with and without the self-energy correction, respectively.}
\label{ehfb1}
\end{figure}
\begin{figure}
\begin{center}
\includegraphics[height=6cm]{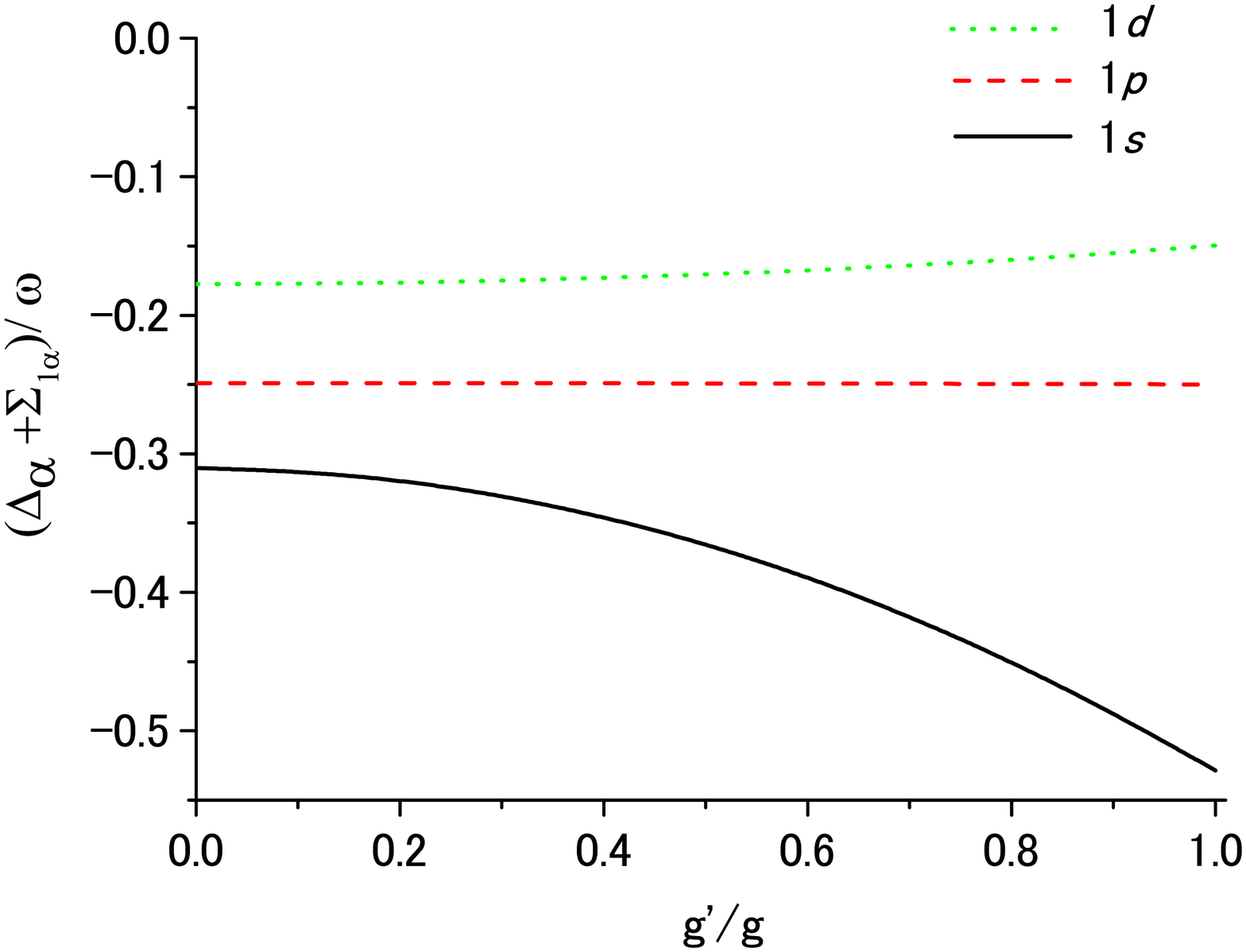}
\end{center}
\caption{(Color online) Pair potential plus the screening term $\Delta_\alpha+\Sigma_{1\alpha}$
as a function of $g'/g$ for $\chi=5$. 
The solid, dashed and dotted lines depict the results for the $1s$, $1p$ and $1d$ states, respectively.
The self-energy is calculated at $\omega_\mu=-\tilde{\epsilon}_\alpha$.}
\label{ehfb4}
\end{figure}

The total energy calculated in ETDHFB (solid line) is shown in Fig. \ref{ehfbtot} as a function of $\chi$,
where $\chi$ is given by $\chi=|g|/\hbar\omega\xi^3$ with
$\xi$ being the oscillator length ($\xi=\sqrt{\hbar/m\omega}$).
In the case of nuclei for which $\hbar\omega\approx 10$ MeV is applied, $C=5$ corresponds to $|g|\approx 400$ MeVfm$^3$, which is
similar to the strength of commonly used pairing interactions for nuclei. 
Both the ETDHFB and HFB results (dotted line) agree well 
with the exact solutions (dot-dashed line).
The better agreement of the ETDHFB results is due to the contribution of the correlation energy as shown in Fig. \ref{ehfbcorr},
where 
the sum $E_{\rm pair}+E_{\rm corr}$ calculated in ETDHFB (solid line) is given as a function of $\chi$. 
In HFB the pairing energy $E_{\rm pair}$ is shown. In the exact case the difference $\Delta E=E_{\rm tot}-E_{\rm MF}$ is shown  (dot-dashed line).
HFB underestimates the correlation energy, which agrees with the results of the pairing model \cite{rich,sandu,lacroix}
and finite nuclei \cite{sandu}.
The deviation of the ETDHFB results from the exact values in Fig. \ref{ehfbcorr} suggests that 
$n_{\alpha\alpha'}$ and ${\rho}_{\alpha\beta\alpha'\beta'}$ in ETDHFB do not completely agree with the exact solutions.
The difference in the total energy is smaller than that in the correlation energy. This is due to a cancellation of errors
between the mean-field energy and the correlation energy \cite{sandu}.

The pairing energy $E_{\rm pair}$ (solid line) and $E_{\rm corr}$ (dot-dashed line) calculated with ETDHFB 
are shown in Fig. \ref{ehfb1} as a function of 
$g'/g$ for $\chi=5$.
The perturbatively calculated correlation energies using 
Eq. (\ref{pertc1}) (the green (gray) dashed line) and Eq. (\ref{pertc}) (green (gray) dot-dashed line) are also shown. The latter has a significant
contribution, which is in agreement with the results for the pairing Hamiltonian \cite{sandu,lacroix}. As mentioned above, the former
describes a correction to the total energy due to the screening effect. In the case of the trapped fermions
it is quite small and plays a role opposite to screening. 
The sum $\Delta_\alpha+\Sigma_{1\alpha}$ is shown in Fig. \ref{ehfb4}
for each single-particle state. The self-energy is calculated at $\omega_\mu=-\tilde{\epsilon}_\alpha$. 
The anti-screening behavior of the correlation energy calculated with 
${\cal C}_{1\alpha\beta\alpha'\beta'}$ is determined by the self-energy $\Sigma_{1\alpha}$ of the $1s$ state.
This indicates that the conditions used to derive Eq. (\ref{F1}) are not fulfilled for the $1s$ state.

We also test the pertubative approximations for the pairing tensor.
The dotted and dashed lines in Fig. \ref{ehfb1} show the results obtained using Eq. (\ref{pertk}) 
with and without the self-energy correction, respectively.
In these calculations the pairing tensor given by Eq. (\ref{pertk})
where $g'$ is used for the higher-order terms (the $v^2$ terms) and the pairing potential in HFB are used in Eq. (\ref{epair}).
Comparison of the results shown by the dotted and dashed lines indicates that
the self-energy correction is significant and almost cancels the screening effect for the pairing tensor. 
This strong cancellation is explained by the facts that the dominant contributions to the sums in Eq. (\ref{pertk}) come
from the $1p$ states because the pairing tensor is the largest for these states and that only the doubly exchanged matrices in the
screening term contribute because of their spin characters of the matrix elements.
As shown in Fig. \ref{ehfb1},
the pairing energy in ETDHFB is slightly increased from the HFB value while the perturbative approach (dotted line) 
gives a slight decrease of the
pairing energy. We found that the coupling to ${\cal C}_{\alpha\beta\alpha'\beta'}$ in $G_{\alpha\beta\gamma:\alpha'}$ 
is responsible for the slight reduction of the pairing correlation in ETDHFB.

\subsection{Tin isotopes}
\begin{figure}
\begin{center}
\includegraphics[height=6cm]{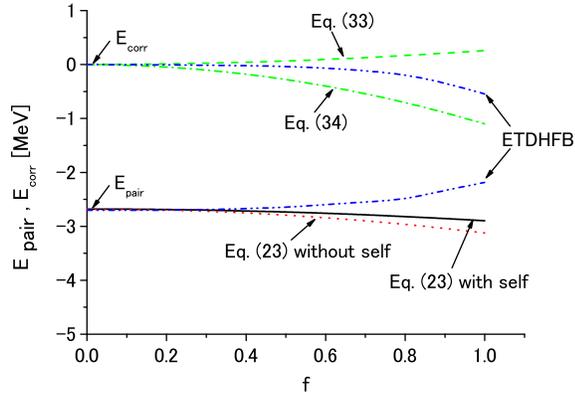}
\end{center}
\caption{(Color online) Pairing energy as a function of $f$ calculated in the perturbative approaches for $^{106}$Sn. 
The solid and dotted lines depict the results with and without the self-energy correction, respectively.
The correlation energy calculated using the perturbative expressions for the two-body correlation matrix
Eq. (\ref{pertc1}) and Eq. (\ref{pertc}) are shown with the green (gray) dashed and dot-dashed lines, respectively.
The pairing energy and correlation energy in ETDHFB are shown with the upper and lower double dot-dashed lines, respectively.}
\label{sn106}
\end{figure}
\begin{figure}
\begin{center}
\includegraphics[height=6cm]{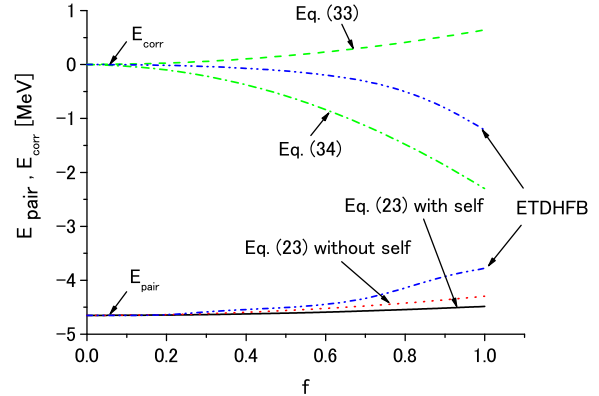}
\end{center}
\caption{(Color online) Same as Fig. \ref{sn106} but for $^{116}$Sn.}
\label{sn116}
\end{figure}
\begin{figure}
\begin{center}
\includegraphics[height=6cm]{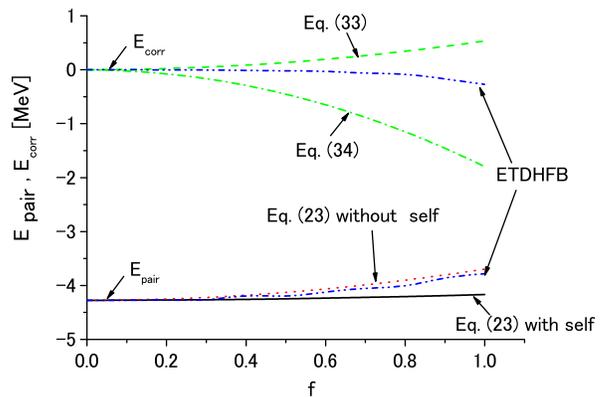}
\end{center}
\caption{(Color online) Same as Fig. \ref{sn106} but for $^{126}$Sn.}
\label{sn126}
\end{figure}
\begin{figure}
\begin{center}
\includegraphics[height=6cm]{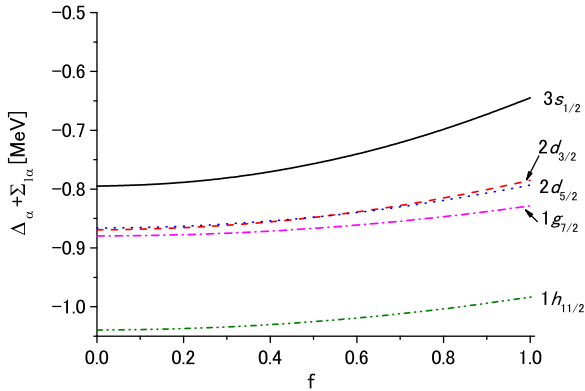}
\end{center}
\caption{(Color online) Sum $\Delta_\alpha+\Sigma_{1\alpha}$
as a function of $f$ for $^{116}$Sn. 
The solid, dashed, dotted, dot-dashed and double dot-dashed lines depict the results for 
the $3s_{1/2}$, $2d_{3/2}$, $2d_{5/2}$, $1g_{7/2}$ and $1h_{11/2}$  states, respectively.
The self-energy is calculated at $\omega_\mu=-\tilde{\epsilon}_\alpha$.}
\label{sndelta}
\end{figure}
In the case of tin isotopes we first perform
the BCS+HF calculations following the numerical procedure used in Ref. \cite{Toh12}. 
The Skyrme III interaction is used to calculate the single-particle states.
For the BCS calculations of $n_{\alpha}$ and 
$\kappa_{\alpha}$ we take the neutron single-particle states,
the $1g_{7/2}$, $2d_{5/2}$, $1h_{11/2}$, $3s_{1/2}$ and $2d_{3/2}$ states. 
As the pairing interaction we use
$v=f_0 (t_0+t_3\rho_p)\delta^3({\bm r}-{\bm r'})$
derived from 
the Skyrme III force with $x_0=0$, where $\rho_p$ is the proton density.
A reduction factor $f_0=0.55$ is used to approximately reproduce the excitation energy of the first $2^+$ state in $^{108}$Sn in extended RPA \cite{Toh12}.
This interaction is similar to a density-dependent pairing interaction 
$v_0(1-\rho/\rho_0)\delta^3({\bm r}-{\bm r'})$, which has often been used in the HFB and quasi-particle RPA calculations.
To simulate the p-h excitations of the core
in the pertubative calculations of the higher-order effects, we add several neutron states in the range
$-20$ MeV $\le \epsilon_\alpha\le 1 $MeV: The continuum states are discretized by confining the wavefunctions in a sphere with radius 15 fm
\cite{Toh12}. There
are two occupied states, ($2p_{1/2}$ and $1g_{9/2}$) and $1\sim4$ unoccupied states 
($2f_{7/2}$, $3p_{1/2}$, $3p_{3/2}$ and $1h_{9/2}$), depending on the isotope. We use the same pairing interaction in the perturbative calculations.

The pairing energies calculated in HF+BCS are $-2.68$ MeV, $-4.65$ MeV and  $-4.27$ MeV for $^{106}$Sn, $^{116}$Sn and $^{126}$Sn, respectively. 
These isotopes correspond to the beginning, middle and end of the subshell.
The pairing energies calculated using the perturbative expression for the pairing tensor Eq. (\ref{pertk}) 
are shown in Figs. \ref{sn106}--\ref{sn126} as a function of the strength $f$ 
of the 
residual interaction: The pairing interaction $v$ used in the second-order terms in Eq. (\ref{pertk}) is multiplied with an artificial factor $f$ ($f=1$ corresponds to the full strength).
As is the case of the trapped fermion system, there is a cancellation between the screening term and the self-energy term.
However, the perturbative correction to the 
pairing tensor is quite small in the case of the tin isotopes.
This may be explained by the fact that the p-h excitation energies in the tin isotopes 
normalized by the averaged pairing potential are a few times larger than those 
in the trapped fermion systems.
The correlation energies calculated using Eq. (\ref{pertc1}) (green (gray) dashed line) 
and Eq. (\ref{pertc}) (green (gray) dot-dashed line) are also shown in Figs. \ref{sn106}--\ref{sn126}. 
The corrections to the total energy from the two-body correlation matrix 
are much larger than those from the pairing tensor. The correlation energies calculated using Eq. (\ref{pertc1}) 
are positive, which means that the pairing correlation is
screened by the process given by the self-energy $\Sigma_{1\alpha}$ as is shown in  Fig. \ref{sndelta}, 
where the sum $\Delta_\alpha+\Sigma_{1\alpha}$ is given
for each single-particle state of $^{116}$Sn. The self-energy is calculated at $\omega_\mu=-\tilde{\epsilon}_\alpha$.
The results shown in Fig \ref{sndelta} indicate that the conditions used in the derivation of Eq. (\ref{F1}) are approximately fulfilled. 

In the ETDHFB calculations we use a small single-particle space
consisting of the neutron $1g_{7/2}$, $2d_{5/2}$, $1h_{11/2}$, $3s_{1/2}$ and $2d_{3/2}$ states 
because it is hard to calculate the two-body matrices using the same single-particle space as used in the perturbative calculations.
The ETDHFB results for the pairing energy (lower double dot-dashed line) and the correlation energy (upper double dot-dashed line) 
are shown in Figs. \ref{sn106}--\ref{sn126} as a function of $f=t/T$, where $T=1200$ fm/c is used. The pairing energies in ETDHFB are slightly increased from the perturbative results,
indicating the contribution of non-perturbative effects as is the case of the trapped fermion system. The correlation energies 
in ETDHF are similar to the sum of the perturbative results from Eqs. (\ref{pertc1}) and (\ref{pertc}) except for $^{126}$Sn.
In the case of $^{126}$Sn the subshell is almost filled and the p-h excitations are limited within the small single-particle space used.

\section{Summary} 
In order to study higher-order effects on the pairing correlation,
we formulated an extended time-dependent Hartree-Fock-Bogoliubov theory (ETDHFB) 
using a truncation scheme of the time-dependent density matrix theory. 
This approach allows us to calculate the pairing tensor and the two-body correlation matrix
in a non-perturbative way and it also is used to
derive their perturbative expressions.
We showed that the perturbative expression for the two-body correlation matrix which contains the pairing tensor 
has a direct connection to other approaches used in the study of the screening effect of the pairing correlation.
We tested ETDHFB for 
fermions trapped in a harmonic potential where comparison
with the exact solution could be made and obtained reasonable agreement with the exact solutions.
We applied the pertubative expressions to
the trapped fermion system and the tin isotopes, and compared with the results in ETDHFB. It was found that for the systems considered, 
the perturbative correction to the pairing energy is small both in the trapped fermion system
and tin isotopes, whereas ETDHFB always gives a slight increase of the pairing energy, indicating the importance of
non-perturbative effects.
It was found that  the perturbative correction to the correlation energy expressed by the pairing tensor shows a screening effect
in the case of the tin isotopes. It was also found that the perturbative corrections to the correlation energy
supplemented by the contribution of two particle - two hole excitations
are similar to the results from full ETDHFB.
The results of our calculations indicate that the screening correction to the results in HFB or BCS+HF 
is at most a few ten percent  
in the case of small finite systems considered here, although more quantitative analysis using larger single-particle space is required.
\appendix
\section{}
We present the terms in the equations of motion for ${\cal C}_{\alpha\beta\alpha'\beta'}$ and $K_{\alpha\beta\gamma:\alpha'}$. 
Since decomposition of higher-order density matrices
to lower-order ones involves various combinations due to the fact that the total wavefunction is not an eigenstate of 
the number operator, these equations
contain many terms. We try to explain the meanings of each term as clearly as possible.

\subsection{} 
The terms in Eq. (\ref{hfbc}) are given below.
$B_{\alpha\beta\alpha'\beta'}$ describes the 2p-2h and 2h-2p excitations as in TDDM \cite{GT}. 
\begin{eqnarray}
B_{\alpha\beta\alpha'\beta'}&=&\sum_{\lambda_1\lambda_2\lambda_3\lambda_4}
\langle\lambda_1\lambda_2|v|\lambda_3\lambda_4\rangle_A
\nonumber \\ 
&\times&[(\delta_{\alpha\lambda_1}-n_{\alpha\lambda_1})(\delta_{\beta\lambda_2}-n_{\beta\lambda_2})
n_{\lambda_3\alpha'}n_{\lambda_4\beta'}
\nonumber \\
&-&n_{\alpha\lambda_1}n_{\beta\lambda_2}(\delta_{\lambda_3\alpha'}-n_{\lambda_3\alpha'})
(\delta_{\lambda_4\beta'}-n_{\lambda_4\beta'})].
\nonumber \\
\end{eqnarray}
Particle - particle and h-h correlations which are not included in the pairing tensor 
are described by $P_{\alpha\beta\alpha'\beta'}$
\begin{eqnarray}
P_{\alpha\beta\alpha'\beta'}&=&\sum_{\lambda_1\lambda_2\lambda_3\lambda_4}
\langle\lambda_1\lambda_2|v|\lambda_3\lambda_4\rangle
\nonumber \\ 
&\times&[(\delta_{\alpha\lambda_1}\delta_{\beta\lambda_2}
-\delta_{\alpha\lambda_1}n_{\beta\lambda_2}
-n_{\alpha\lambda_1}\delta_{\beta\lambda_2})
{\cal C}_{\lambda_3\lambda_4\alpha'\beta'}
\nonumber \\
&-&(\delta_{\lambda_3\alpha'}\delta_{\lambda_4\beta'}
-\delta_{\lambda_3\alpha'}n_{\lambda_4\beta'}
-n_{\lambda_3\alpha'}\delta_{\lambda_4\beta'})
{\cal C}_{\alpha\beta\lambda_1\lambda_2}].
\nonumber \\
\end{eqnarray}
$H_{\alpha\beta\alpha'\beta'}$ describes p-h correlations.
\begin{eqnarray}
H_{\alpha\beta\alpha'\beta'}&=&\sum_{\lambda_1\lambda_2\lambda_3\lambda_4}
\langle\lambda_1\lambda_2|v|\lambda_3\lambda_4\rangle_A
\nonumber \\ 
&\times&[\delta_{\alpha\lambda_1}(n_{\lambda_3\alpha'}{\cal C}_{\lambda_4\beta\lambda_2\beta'}
-n_{\lambda_3\beta'}{\cal C}_{\lambda_4\beta\lambda_2\alpha'})
\nonumber \\
&+&\delta_{\beta\lambda_2}(n_{\lambda_4\beta'}{\cal C}_{\lambda_3\alpha\lambda_1\alpha'}
-n_{\lambda_4\alpha'}{\cal C}_{\lambda_3\alpha\lambda_1\beta'})
\nonumber \\
&-&\delta_{\alpha'\lambda_3}(n_{\alpha\lambda_1}{\cal C}_{\lambda_4\beta\lambda_2\beta'}
-n_{\beta\lambda_1}{\cal C}_{\lambda_4\alpha\lambda_2\beta'})
\nonumber \\
&-&\delta_{\beta'\lambda_4}(n_{\beta\lambda_2}{\cal C}_{\lambda_3\alpha\lambda_1\alpha'}
-n_{\alpha\lambda_2}{\cal C}_{\lambda_3\beta\lambda_1\alpha'})].
\nonumber \\
\label{h-term}
\end{eqnarray}
The coupling to the pairing tensor is given by $S_{\alpha\beta\alpha'\beta'}$.
\begin{eqnarray}
S_{\alpha\beta\alpha'\beta'}&=&\sum_{\lambda_1\lambda_2\lambda_3\lambda_4}
\langle\lambda_1\lambda_2|v|\lambda_3\lambda_4\rangle_A
\nonumber \\ 
&\times&[\delta_{\alpha\lambda_1}(n_{\lambda_3\alpha'}\kappa_{\lambda_4\beta}\kappa^*_{\lambda_2\beta'}
-n_{\lambda_3\beta'}\kappa_{\lambda_4\beta}\kappa^*_{\lambda_2\alpha'})
\nonumber \\
&+&\delta_{\beta\lambda_2}(n_{\lambda_4\beta'}\kappa_{\lambda_3\alpha}\kappa^*_{\lambda_1\alpha'}
-n_{\lambda_4\alpha'}\kappa_{\lambda_3\alpha}\kappa^*_{\lambda_1\beta'})
\nonumber \\
&-&\delta_{\alpha'\lambda_3}(n_{\alpha\lambda_1}\kappa_{\lambda_4\beta}\kappa^*_{\lambda_2\beta'}
-n_{\beta\lambda_1}\kappa_{\lambda_4\alpha}\kappa^*_{\lambda_2\beta'})
\nonumber \\
&-&\delta_{\beta'\lambda_4}(n_{\beta\lambda_2}\kappa_{\lambda_3\alpha}\kappa^*_{\lambda_1\alpha'}
-n_{\alpha\lambda_2}\kappa_{\lambda_3\beta}\kappa^*_{\lambda_1\alpha'})].
\nonumber \\
\label{s-term}
\end{eqnarray}
From the decomposition
\begin{eqnarray}
&\langle\Phi(t)|&a^\dag_{\alpha'}a^\dag_{\beta'}a^\dag_{\gamma'}a_{\gamma}a_{\beta}a_{\alpha}|\Phi(t)\rangle
\nonumber \\
&=&
\langle\Phi(t)|a^\dag_{\alpha'}a^\dag_{\beta'}|\Phi(t)\rangle\langle\Phi(t)|a^\dag_{\gamma'}a_{\gamma}a_{\beta}a_{\alpha}|\Phi(t)\rangle
\nonumber \\
&+&\cdot~\cdot
\end{eqnarray}
we obtain $T_{\alpha\beta\alpha'\beta'}$ which expresses the coupling to $K_{\alpha\beta\gamma:\alpha'}$:
\begin{eqnarray}
T_{\alpha\beta\alpha'\beta'}&=&\sum_{\lambda}
(\Delta_{\alpha\lambda}K^*_{\lambda\beta'\alpha':\beta}-\Delta_{\beta\lambda}K^*_{\lambda\beta'\alpha':\alpha}
\nonumber \\ 
&-&\Delta^*_{\alpha'\lambda}K_{\alpha\beta\lambda:\beta'}+\Delta^*_{\beta'\lambda}K_{\alpha\beta\lambda:\alpha'})
\nonumber \\
&+&\frac{1}{2}\sum_{\lambda_1\lambda_2\lambda_3\lambda_4}\langle\lambda_1\lambda_2|v|\lambda_3\lambda_4\rangle_A
\nonumber \\ 
&\times&[\delta_{\alpha\lambda_1}(2\kappa_{\beta\lambda_4}K^*_{\beta'\lambda_2\alpha':\lambda_3}
+\kappa^*_{\beta'\lambda_2}K_{\beta\lambda_4\lambda_3:\alpha'}
\nonumber \\
&-&\kappa^*_{\alpha'\lambda_2}K_{\beta\lambda_4\lambda_3:\beta'})
\nonumber \\
&-&\delta_{\beta\lambda_1}(2\kappa_{\alpha\lambda_4}K^*_{\beta'\lambda_2\alpha':\lambda_3}
+\kappa^*_{\beta'\lambda_2}K_{\alpha\lambda_4\lambda_3:\alpha'}
\nonumber \\
&-&\kappa^*_{\alpha'\lambda_2}K_{\alpha\lambda_4\lambda_3:\beta'})
\nonumber \\
&-&\delta_{\alpha'\lambda_3}(2\kappa^*_{\lambda_2\beta'}K_{\alpha\lambda_4\beta:\lambda_1}
+\kappa_{\alpha\lambda_4}K^*_{\lambda_1\lambda_2\beta':\beta}
\nonumber \\
&-&\kappa_{\beta\lambda_4}K^*_{\lambda_1\lambda_2\beta':\alpha})
\nonumber \\
&+&\delta_{\beta'\lambda_3}(2\kappa^*_{\lambda_2\alpha'}K_{\alpha\lambda_4\beta:\lambda_1}
+\kappa_{\alpha\lambda_4}K^*_{\lambda_1\lambda_2\alpha':\beta}
\nonumber \\
&-&\kappa_{\beta\lambda_4}K^*_{\lambda_1\lambda_2\alpha':\alpha})].
\label{t-term}
\end{eqnarray}
The terms in the first sum describe the coupling to the pairing potential. Since the terms in the second sum
contain both p-p (and h-h) and p-h correlations, they may describe corrections to $P_{\alpha\beta\alpha'\beta'}$ and $H_{\alpha\beta\alpha'\beta'}$. 
In the derivation of  Eq. (\ref{hfbc}) we neglected the genuine three-body density matrix $\langle\Phi(t)|a^\dag_{\alpha'}a^\dag_{\beta'}a^\dag_{\gamma'}a_{\gamma}a_{\beta}a_{\alpha}|\Phi(t)\rangle$
as in TDDM.

\subsection{}
The terms in Eq. (\ref{hfb3}) are given below.
$D_{\alpha\beta\gamma:\alpha'}$ describes the coupling to one pairing tensor
\begin{eqnarray}
D_{\alpha\beta\gamma:\alpha'}&=&-\sum_{\lambda_1\lambda_2}{(\langle\alpha\beta|v|\lambda_1\lambda_2\rangle_A{\kappa_{\gamma\lambda_2}}+\langle\beta\gamma|v|\lambda_1\lambda_2\rangle_A{\kappa_{\alpha\lambda_2}}}
\nonumber \\
&-&{\langle\alpha\gamma|v|\lambda_1\lambda_2\rangle_A{\kappa_{\beta\lambda_2}}){n_{\lambda_1\alpha'}}}
\nonumber \\
&+&\sum_{\lambda_1\lambda_2\lambda_3}[\langle\alpha\lambda_1|v|\lambda_2\lambda_3\rangle_A(n_{\beta\lambda_1}{\kappa_{\gamma\lambda_3}}
-n_{\gamma\lambda_1}\kappa_{\beta\lambda_3})
\nonumber \\
&+&\langle\beta\lambda_1|v|\lambda_2\lambda_3\rangle_A(n_{\gamma\lambda_1}{\kappa_{\alpha\lambda_3}}
-n_{\alpha\lambda_1}\kappa_{\gamma\lambda_3})
\nonumber \\
&+&\langle\gamma\lambda_1|v|\lambda_2\lambda_3\rangle_A(n_{\alpha\lambda_1}{\kappa_{\beta\lambda_3}}
-n_{\beta\lambda_1}\kappa_{\alpha\lambda_3})]{n_{\lambda_2\alpha'}}
\nonumber \\
&+&\sum_{\lambda_1\lambda_2\lambda_3}\langle\lambda_1\lambda_2|v|\alpha'\lambda_3\rangle_A
\nonumber \\ 
&\times&(n_{\alpha\lambda_1}n_{\gamma\lambda_2}\kappa_{\beta\lambda_3}
-n_{\beta\lambda_1}n_{\gamma\lambda_2}\kappa_{\alpha\lambda_3}
\nonumber \\
&-&n_{\alpha\lambda_1}n_{\beta\lambda_2}\kappa_{\gamma\lambda_3}).
\label{d-term}
\end{eqnarray}
The terms in the first sum originate from the decomposition 
\begin{eqnarray}
&\langle\Phi(t)|&a^\dag_{\alpha'}a_{\gamma}a_{\beta}a_{\alpha}|\Phi(t)\rangle
\nonumber \\
&=&\langle\Phi(t)|a^\dag_{\alpha'}a_{\gamma}|\Phi(t)\rangle\langle\Phi(t)|a_{\beta}a_{\alpha}|\Phi(t)\rangle
\nonumber \\
&+&\cdot~\cdot,
\end{eqnarray}
whereas those in the second and third sums from
\begin{eqnarray}
&\langle\Phi(t)|&a^\dag_{\alpha'}a^\dag_{\beta'}a_{\delta}a_{\gamma}a_{\beta}a_{\alpha}|\Phi(t)\rangle
\nonumber \\
&=&
\langle\Phi(t)|a^\dag_{\alpha'}a_{\alpha}|\Phi(t)\rangle\langle\Phi(t)|a^\dag_{\beta'}a_{\delta}|\Phi(t)\rangle 
\nonumber \\
&\times&\langle\Phi(t)|a_{\gamma}a_{\beta}|\Phi(t)\rangle
+\cdot~\cdot.
\end{eqnarray}
The perturbative expression for the pairing tensor Eq. (\ref{pertk}) is obtained from
the first term and $D_{\alpha\beta\gamma:\alpha'}$ in Eq. (\ref{hfb3}). From the decomposition of the matrix
\begin{eqnarray}
&\langle\Phi(t)|&a^\dag_{\alpha'}a^\dag_{\beta'}a_{\delta}a_{\gamma}a_{\beta}a_{\alpha}|\Phi(t)\rangle
\nonumber \\
&=&
\langle\Phi(t)|a^\dag_{\alpha'}a^\dag_{\beta'}|\Phi(t)\rangle\langle\Phi(t)|a_{\delta}a_{\gamma}|\Phi(t)\rangle
\nonumber \\
&\times&\langle\Phi(t)|a_{\beta}a_{\alpha}|\Phi(t)\rangle
%\nonumber \\
+\cdot~\cdot,
\end{eqnarray} 
we also obtain
the coupling to three paring tensors given by $E_{\alpha\beta\gamma:\alpha'}$, 
\begin{eqnarray}
E_{\alpha\beta\gamma:\alpha'}&=&-\sum_{\lambda_1\lambda_2\lambda_3}(\langle\alpha\lambda_1|v|\lambda_2\lambda_3\rangle_A\kappa_{\beta\lambda_2}\kappa_{\gamma\lambda_3}
\nonumber \\
&-&\langle\beta\lambda_1|v|\lambda_2\lambda_3\rangle_A \kappa_{\alpha\lambda_2}\kappa_{\gamma\lambda_3}
\nonumber \\ 
&+&\langle\gamma\lambda_1|v|\lambda_2\lambda_3\rangle_A\kappa_{\alpha\lambda_2}\kappa_{\beta\lambda_3})\kappa^*_{\alpha'\lambda_1}.
\label{e-term}
\end{eqnarray}
These terms express the modification of the two-particle propagator due to the pairing correlations with other particles.
The terms in $F_{\alpha\beta\gamma:\alpha'}$ are from 
\begin{eqnarray}
&\langle\Phi(t)|&a^\dag_{\alpha'}a^\dag_{\beta'}a_{\delta}a_{\gamma}a_{\beta}a_{\alpha}|\Phi(t)\rangle
\nonumber \\
&=&
\langle\Phi(t)|a^\dag_{\alpha'}a_{\alpha}|\Phi(t)\rangle\langle\Phi(t)|a^\dag_{\beta'}a_{\delta}a_{\gamma}a_{\beta}|\Phi(t)\rangle
\nonumber \\
&+&\cdot~\cdot
\end{eqnarray}
and describe correlations among $K_{\alpha\beta\gamma:\alpha'}$:
\begin{eqnarray}
F_{\alpha\beta\gamma:\alpha'}&=&\frac{1}{2}\sum_{\lambda_1\lambda_2\lambda_3\lambda_4}\langle\lambda_1\lambda_2|v|\lambda_3\lambda_4\rangle_A
\nonumber \\ 
&\times&[(\delta_{\alpha\lambda_1}\delta_{\beta\lambda_2}-\delta_{\alpha\lambda_1}n_{\beta\lambda_2}-\delta_{\beta\lambda_2}n_{\alpha\lambda_1})K_{\gamma\lambda_3\lambda_4:\alpha'}
\nonumber \\
&+&(\delta_{\beta\lambda_1}\delta_{\gamma\lambda_2}-\delta_{\beta\lambda_1}n_{\gamma\lambda_2}-\delta_{\gamma\lambda_2}n_{\beta\lambda_1})K_{\alpha\lambda_3\lambda_4:\alpha'}
\nonumber \\
&-&(\delta_{\alpha\lambda_1}\delta_{\gamma\lambda_2}-\delta_{\alpha\lambda_1}n_{\gamma\lambda_2}-\delta_{\gamma\lambda_2}n_{\alpha\lambda_1})K_{\beta\lambda_3\lambda_4:\alpha'}]
\nonumber \\
&+&\sum_{\lambda_1\lambda_2\lambda_3\lambda_4}\langle\lambda_1\lambda_2|v|\lambda_3\lambda_4\rangle_A
\nonumber \\
&\times&[\frac{1}{2}(\delta_{\alpha\lambda_1}K_{\beta\gamma\lambda_3:\lambda_2}-\delta_{\beta\lambda_1}K_{\alpha\gamma\lambda_3:\lambda_2}
\nonumber \\
&+&\delta_{\gamma\lambda_1}K_{\alpha\beta\lambda_3:\lambda_2})n_{\lambda_4\alpha'}
\nonumber \\
%&+&\sum_{\lambda_1\lambda_2\lambda_3}\langle\lambda_1\lambda_2|v|\alpha'\lambda_3\rangle_A
%\nonumber \\ 
&+&\delta_{\lambda_3\alpha'}(n_{\gamma\lambda_2}K_{\alpha\beta\lambda_4:\lambda_1}-n_{\beta\lambda_2}K_{\alpha\gamma\lambda_4:\lambda_1}
\nonumber \\
&-&n_{\alpha\lambda_2}K_{\gamma\beta\lambda_4:\lambda_1})].
\end{eqnarray}
The terms in the first sum describe p-p correlations while those in the second sum p-h correlations.
The terms in $G_{\alpha\beta\gamma:\alpha'}$ come from 
\begin{eqnarray}
&\langle\Phi(t)|&a^\dag_{\alpha'}a^\dag_{\beta'}a_{\delta}a_{\gamma}a_{\beta}a_{\alpha}|\Phi(t)\rangle
\nonumber \\
&=&
\langle\Phi(t)|a_{\delta}a_{\gamma}|\Phi(t)\rangle\langle\Phi(t)|a^\dag_{\alpha'}a^\dag_{\beta'}a_{\beta}a_{\alpha}|\Phi(t)\rangle
\nonumber \\
&+&\cdot~\cdot:
\end{eqnarray}
%\newpage
\begin{eqnarray}
G_{\alpha\beta\gamma:\alpha'}&=&
\sum_{\lambda}{(\Delta_{\alpha\lambda}{\cal C}_{\beta\gamma\alpha'\lambda}-\Delta_{\beta\lambda}{\cal C}_{\alpha\gamma\alpha'\lambda}+\Delta_{\gamma\lambda}{\cal C}_{\alpha\beta\alpha'\lambda})}
\nonumber \\
&-&\sum_{\lambda_1\lambda_2\lambda_3\lambda_4}\langle\lambda_1\lambda_2|v|\lambda_3\lambda_4\rangle_A
\nonumber \\ 
&\times&[(\delta_{\alpha\lambda_1}\kappa_{\beta\lambda_3}-\delta_{\beta\lambda_1}\kappa_{\alpha\lambda_3}){\cal C}_{\gamma\lambda_4\alpha'\lambda_2})
\nonumber \\
&+&(\delta_{\beta\lambda_1}\kappa_{\gamma\lambda_3}-\delta_{\gamma\lambda_1}\kappa_{\beta\lambda_3}){\cal C}_{\alpha\lambda_4\alpha'\lambda_2})
\nonumber \\
&-&(\delta_{\alpha\lambda_1}\kappa_{\gamma\lambda_3}-\delta_{\gamma\lambda_1}\kappa_{\alpha\lambda_3}){\cal C}_{\beta\lambda_4\alpha'\lambda_2})].
\label{g-term}
\end{eqnarray}
These terms describe the coupling to ${\cal C}_{\alpha\beta\alpha'\beta'}$.
In the above derivation of Eq. (\ref{hfb3}) the genuine correlated matrices $\langle\Phi(t)|a^\dag_{\alpha'}a^\dag_{\beta'}a_{\delta}a_{\gamma}a_{\beta}a_{\alpha}|\Phi(t)\rangle$
and $\langle\Phi(t)|a_{\delta}a_{\gamma}a_{\beta}a_{\alpha}|\Phi(t)\rangle$ are neglected.

\section{}
We consider the Gorkov Green's function
\begin{eqnarray}
{\cal G}_{\alpha\beta}(t,t')=
\left(
\begin{array}{cc}
G_{\alpha\beta}(t,t')&F_{\alpha\beta}(t,t')\\
-F^*_{\alpha\beta}(t,t')&-G^*_{\alpha\beta}(t,t')
\end{array}
\right),
\end{eqnarray}
where $iG_{\alpha\beta}(t,t')=\langle0|T(a_\alpha(t)a^\dag_\beta(t'))|0\rangle$ and $iF_{\alpha\beta}(t,t')=\langle 0|T(a_\alpha(t)a_\beta(t'))|0\rangle$ 
with $a_\alpha(t)=\exp[i(H-\mu\hat{N})t/\hbar]a_\alpha\exp[-i(H-\mu\hat{N})t/\hbar]$. 
The Green's functions are written in terms of the transition amplitudes
$x^\mu_\alpha=\langle\mu|a_\alpha|0\rangle$ and $y^\mu_\alpha=\langle\mu|a^\dag_\alpha|0\rangle$ as 
\begin{eqnarray}
iG_{\alpha\beta}(t,t')&=&\theta(t-t')\langle0|a_\alpha(t)a^\dag_\beta(t')|0\rangle
\nonumber \\
&-&\theta(t'-t)\langle 0|a^\dag_\beta(t')a_\alpha(t)|0\rangle
\nonumber \\
&=&\sum_\mu[\theta(t-t')\langle0|a_\alpha|\mu\rangle\langle\mu|a^\dag_\beta|0\rangle
\nonumber \\
&\times&e^{-i\omega_\mu(t-t')/\hbar}
\nonumber \\
&-&\theta(t'-t)\langle 0|a^\dag_\beta|\mu\rangle\langle\mu|a_\alpha|0\rangle
\nonumber \\
&\times&e^{-i\omega_\mu(t'-t)/\hbar}]
\nonumber \\
&=&\sum_\mu[\theta(t-t')(y_\alpha^\mu)^*y_\beta^\mu e^{-i\omega_\mu(t-t')/\hbar}
\nonumber \\
&-&\theta(t'-t)(x_\beta^\mu)^*x_\alpha^\mu e^{-i\omega_\mu(t'-t)/\hbar}],
\end{eqnarray}
\begin{eqnarray}
iF_{\alpha\beta}(t,t')&=&\theta(t-t')\langle0|a_\alpha(t)a_\beta(t')|0\rangle
\nonumber \\
&-&\theta(t'-t)\langle 0|a_\beta(t')a_\alpha(t)|0\rangle
\nonumber \\
&=&\sum_\mu[\theta(t-t')\langle0|a_\alpha|\mu\rangle\langle\mu|a_\beta|0\rangle
\nonumber \\
&\times&e^{-i\omega_\mu(t-t')/\hbar}
\nonumber \\
&-&\theta(t'-t)\langle 0|a_\beta|\mu\rangle\langle\mu|a_\alpha|0\rangle
\nonumber \\
&\times&e^{-i\omega_\mu(t'-t)/\hbar}]
\nonumber \\
&=&\sum_\mu[\theta(t-t')(y_\alpha^\mu)^*x_\beta^\mu e^{-i\omega_\mu(t-t')/\hbar}
\nonumber \\
&-&\theta(t'-t)(y_\beta^\mu)^*x_\alpha^\mu e^{-i\omega_\mu(t'-t)/\hbar}].
\end{eqnarray}
The equations of motion for the Green's functions can be formulated using 
the equations of motion for the transition amplitudes $x^\mu_\alpha$ and $y^\mu_\alpha$ \cite{soma}.
First we derive the perturbative expressions for the self-energies of the Green's function $G_{\alpha\beta}(t,t')$ which are related to corrections to the pairing potential
and the mean-field potential.
The equation motion for $x^\mu_\alpha$ is written as 
\begin{eqnarray}
\omega_\mu x^\mu_\alpha&=&\langle\mu|[H-\mu\hat{N},a_\alpha]|0\rangle\
\nonumber \\
&=&-\tilde{\epsilon}_{\alpha}x^\mu_\alpha-\Delta_{\alpha}y^\mu_{\bar{\alpha}}
\nonumber \\
&-&\frac{1}{2}\sum_{\lambda_1\lambda_2\lambda_3}\langle\alpha\lambda_1|v|\lambda_2\lambda_3\rangle_A X^\mu_{\lambda_2\lambda_3:\lambda_1},
\label{x1a}
\end{eqnarray}
where
$X^\mu_{\alpha\beta:\alpha'}=\langle\mu|a^\dag_{\alpha'}a_{\beta}a_{\alpha}|0\rangle$.
We assume that $\epsilon_{\alpha\alpha'}=\epsilon_\alpha \delta_{\alpha\alpha'}$, $n_{\alpha\alpha'}=n_\alpha \delta_{\alpha\alpha'}$ and $\Delta_{\alpha\beta}=\Delta_\alpha\delta_{\beta\bar{\alpha}}$.
The equation of motion for $X^\mu_{\alpha\beta:\alpha'}$ contains the terms proportional to $y^\mu_\alpha$ and $x^\mu_\alpha$ 
\begin{eqnarray}
\omega_\mu X^\mu_{\alpha\beta:\alpha'}&=&\langle\mu|[H-\mu\hat{N},a^\dag_{\alpha'}a_{\beta}a_{\alpha}]|0\rangle
\nonumber \\
&=&(\tilde{\epsilon}_{\alpha'}-\tilde{\epsilon}_{\alpha}-\tilde{\epsilon}_{\beta})X^\mu_{\alpha\beta:\alpha'}
\nonumber \\
&+&\sum_{\lambda}[\langle\lambda\alpha|v|\alpha'\bar{\beta}\rangle_A(n_\alpha-n_{\alpha'})\kappa_\beta
\nonumber \\
&-&\langle\lambda\beta|v|\alpha'\bar{\alpha}\rangle_A(n_\beta-n_{\alpha'})\kappa_\alpha
]y^\mu_{\lambda}
\nonumber \\
&+&\sum_{\lambda}\langle\alpha\beta|v|\alpha'\lambda\rangle_A
(\bar{n}_\alpha\bar{n}_\beta n_{\alpha'}+n_\alpha n_\beta\bar{n}_{\alpha'})x^\mu_\lambda
\nonumber \\
&+&{\rm more~terms~with}~X^\mu_{\alpha\beta:\alpha'}.
\label{x2a}
\end{eqnarray}
Inserting $X^\mu_{\alpha\beta:\alpha'}$ into Eq. (\ref{x1a}), we obtain 
\begin{eqnarray}
\omega_\mu x^\mu_\alpha&=&
-\tilde{\epsilon}_{\alpha}x^\mu_\alpha-\Delta_{\alpha}y^\mu_{\bar{\alpha}}
\nonumber \\
&-&\sum_{\lambda\lambda_1\lambda_2\lambda_3}[
\langle\alpha\lambda_1|v|\lambda_2\lambda_3\rangle_A
\frac{n_{\lambda_3}-n_{\lambda_1}}{\omega_\mu+\tilde{\epsilon}_{\lambda_2}+\tilde{\epsilon}_{\lambda_3}-\tilde{\epsilon}_{\lambda_1}}
\nonumber \\
&\times&\langle\lambda\lambda_3|v|\bar{\lambda}_2\lambda_1\rangle_A\kappa_{\lambda_2}]y^\mu_\lambda
\nonumber \\
&+&\frac{1}{2}\sum_{\lambda\lambda_1\lambda_2\lambda_3}[
\langle\alpha\lambda_1|v|\lambda_2\lambda_3\rangle_A
\frac{\bar{n}_{\lambda_2}\bar{n}_{\lambda_3}n_{\lambda_1}+n_{\lambda_2}n_{\lambda_3}\bar{n}_{\lambda_1}}
{\omega_\mu+\tilde{\epsilon}_{\lambda_2}+\tilde{\epsilon}_{\lambda_3}-\tilde{\epsilon}_{\lambda_1}}
\nonumber \\
&\times&\langle\lambda_2\lambda_3|v|\lambda\lambda_1\rangle_A]x^\mu_\lambda.
\label{x11a}
\end{eqnarray}
The third term is the perturbative expression of the self-energy 
describing a correction to the pairing potential $\Delta_\alpha$
and the last term a correction to the mean-field potential. The diagonal part of the third term $\Sigma_{1\alpha}$ is given as 
\begin{eqnarray}
\Sigma_{1\alpha}&=&
\sum_{\lambda\lambda_1\lambda_2\lambda_3}
\langle\alpha\lambda_1|v|\lambda_2\lambda_3\rangle_A
\nonumber \\
&\times&
\frac{n_{\lambda_3}-n_{\lambda_1}}{\omega_\mu+\tilde{\epsilon}_{\lambda_2}+\tilde{\epsilon}_{\lambda_3}-\tilde{\epsilon}_{\lambda_1}}
%\nonumber \\
%&\times&
\langle\bar{\alpha}\lambda_3|v|\bar{\lambda}_2\lambda_1\rangle_A\kappa_{\lambda_2}.
\label{x12a}
\end{eqnarray}
Similarly, the self-energy $\Sigma_{2\alpha}$ for the last term of Eq. (\ref{x11a}) is given by
\begin{eqnarray}
\Sigma_{2\alpha}&=&-\frac{1}{2}\sum_{\lambda\lambda_1\lambda_2\lambda_3}
\langle\alpha\lambda_1|v|\lambda_2\lambda_3\rangle_A
\frac{\bar{n}_{\lambda_2}\bar{n}_{\lambda_3}n_{\lambda_1}+n_{\lambda_2}n_{\lambda_3}\bar{n}_{\lambda_1}}
{\omega_\mu+\tilde{\epsilon}_{\lambda_2}+\tilde{\epsilon}_{\lambda_3}-\tilde{\epsilon}_{\lambda_1}}
\nonumber \\
&\times&\langle\lambda_2\lambda_3|v|\alpha\lambda_1\rangle_A.
\label{x12b}
\end{eqnarray}

Next we show that the equation for the pairing tensor (Eq. (\ref{hfb2})) is derived from that for $F_{\alpha\beta}(t,t')$.
This is because the pairing tensor is given as the equal-time limit of 
$F_{\alpha\beta}(t,t')$ as 
\begin{eqnarray}
\lim_{t'\rightarrow t+0}(-i)F_{\alpha\beta}(t,t')=\kappa_{\alpha\beta}=\sum_\mu(y_\beta^\mu)^*x_\alpha^\mu.
\end{eqnarray}
The equation motion for $y^\mu_\alpha$ is written as 
\begin{eqnarray}
\omega_\mu y^\mu_\alpha&=&\langle\mu|[H-\mu\hat{N},a^\dag_\alpha]|0\rangle\
\nonumber \\
&=&\tilde{\epsilon}_{\alpha}y^\mu_\alpha+\Delta_{\alpha}^*x^\mu_{\bar{\alpha}}
\nonumber \\
&+&\frac{1}{2}\sum_{\lambda_1\lambda_2\lambda_3}\langle\lambda_1\lambda_2|v|\alpha\lambda_3\rangle_A Y^\mu_{\lambda_3:\lambda_1\lambda_2},
\label{y1a}
\end{eqnarray}
where
$Y^\mu_{\alpha':\alpha\beta}=\langle\mu|a^\dag_{\alpha}a^\dag_{\beta}a_{\alpha'}|0\rangle$.
Using Eq. (\ref{x1a}) and the complex conjugate of Eq. (\ref{y1a}) (we assume $\omega_\mu$ is real), we calculate 
$\sum_{\mu}[\omega_\mu(y_{\bar{\alpha}}^\mu)^*x^\mu_\alpha-(y^\mu_{\bar{\alpha}})^*\omega_\mu x^\mu_\alpha]$ 
and obtain
\begin{eqnarray}
0&=&2\tilde{\epsilon}_\alpha\sum_\mu (y^\mu_{\bar{\alpha}})^*x^\mu_\alpha
+\Delta_\alpha\sum_\mu (y^\mu_{\bar{\alpha}})^*y^\mu_{\bar{\alpha}}
\nonumber \\
&-&\Delta_\alpha\sum_\mu (x^\mu_{\alpha})^*x^\mu_{\alpha}
+\frac{1}{2}\sum_{\lambda_1\lambda_2\lambda_3}\langle\alpha\lambda_1|v|\lambda_2\lambda_3\rangle_A
\nonumber \\
&\times&\sum_\mu(y^\mu_{\bar{\alpha}})^*X^\mu_{\lambda_2\lambda_3:\lambda_1}
\nonumber \\
&+&\frac{1}{2}\sum_{\lambda_1\lambda_2\lambda_3}\langle\bar{\alpha}\lambda_1|v|\lambda_2\lambda_3\rangle_A
\nonumber \\
&\times&\sum_\mu(Y^\mu_{\lambda_1:\lambda_2\lambda_3})^*x^\mu_{{\alpha}}.
\label{yx}
\end{eqnarray}
When the following replacements, $\sum_\mu (y^\mu_{\bar{\alpha}})^*x^\mu_\alpha=\kappa_\alpha$, $\sum_\mu (y^\mu_{\bar{\alpha}})^*y^\mu_{\bar{\alpha}}=1-n_\alpha$,
$\sum_\mu (x^\mu_{\alpha})^*x^\mu_{\alpha}=n_\alpha$, and $\sum_\mu(y^\mu_{\bar{\alpha}})^*X^\mu_{\lambda_2\lambda_3:\lambda_1}=
-\sum_\mu(Y^\mu_{\lambda_1:\lambda_2\lambda_3})^*x^\mu_\alpha=-K_{\alpha\lambda_2\lambda_3:\lambda_1}$ are made,
the above equation is of the same form as Eq. (\ref{hfb2}) for a stationary solution.
From the equations of motion for $x^\mu_\alpha$, $y^\mu_{{\alpha}}$, $X^\mu_{\alpha\beta:\alpha'}$ and $Y^\mu_{\alpha':\alpha\beta}$,
we can derive the pertubative expression for $\kappa_{\alpha}$ (Eq. (\ref{pertk0})). Let us discuss this point in some more detail.
Considering $\sum_\mu[\omega_\mu(y_\gamma^\mu)^*X_{\alpha\beta:\alpha'}^\mu-(y_\gamma^\mu)^*\omega_\mu X_{\alpha\beta:\alpha'}^\mu]$,
we show that the term $\sum_\mu(y_\gamma^\mu)^*X_{\alpha\beta:\alpha}^\mu$ on the right-hand side of Eq. (\ref{yx}) is reduced to 
$-K_{\alpha\beta\gamma:\alpha'}$ given in Eq. (\ref{K}).
From the equations of motion for $y_\gamma^\mu$ and $X_{\alpha\beta:\alpha'}^\mu$ we obtain
\begin{eqnarray}
0&=&(\tilde{\epsilon}_\alpha+\tilde{\epsilon}_\beta+\tilde{\epsilon}_\gamma-\tilde{\epsilon}_{\alpha'})\sum_\mu(y_\gamma^\mu)^*X_{\alpha\beta:\alpha'}^\mu
\nonumber \\
&-&\sum_{\mu\lambda}[\langle\lambda\alpha|v|\alpha'\bar{\beta}\rangle_A(n_\alpha-n_{\alpha'})\kappa_\beta
\nonumber \\
&-&\langle\lambda\beta|v|\alpha'\bar{\alpha}\rangle_A(n_\beta-n_{\alpha'})\kappa_\alpha
](y_\gamma^\mu)^*y^\mu_{\lambda}
\nonumber \\
&-&\sum_{\mu\lambda}\langle\alpha\beta|v|\alpha'\lambda\rangle_A
(\bar{n}_\alpha\bar{n}_\beta n_{\alpha'}+n_\alpha n_\beta\bar{n}_{\alpha'})(y_\gamma^\mu)^*x^\mu_\lambda
\nonumber \\
&+&\frac{1}{2}\sum_{\mu\lambda_1\lambda_2\lambda_3}\langle\gamma\lambda_3|v|\lambda_1\lambda_2\rangle_A (Y^\mu_{\lambda_3:\lambda_1\lambda_2})^*X_{\alpha\beta:\alpha'}^\mu.
\label{K1}
\end{eqnarray}
If we use $\sum_\mu (y^\mu_{\beta})^*x^\mu_\alpha=\delta_{\beta\bar{\alpha}}\kappa_\alpha$, $\sum_\mu (y^\mu_{\alpha})^*y^\mu_{\beta}=\delta_{\alpha\beta}(1-n_\alpha)$ 
and the additional relation 
\begin{eqnarray}
\sum_\mu (Y^\mu_{\alpha':\alpha\beta})^*X^\mu_{\sigma\rho:\sigma'}&=&\sum_\mu\langle0|a^\dag_{\alpha'}a_{\beta}a_{\alpha}|\mu\rangle
\langle\mu|a^\dag_{\sigma'}a_{\rho}a_{\sigma}|0\rangle
\nonumber \\
&\approx&
\delta_{\sigma'\alpha}(\delta_{\sigma\bar{\beta}}\delta_{\rho\alpha'}-\delta_{\rho\bar{\beta}}\delta_{\sigma\alpha'})\bar{n}_\alpha n_{\alpha'}\kappa_\beta
\nonumber \\
&-&\delta_{\sigma'\beta}(\delta_{\sigma\bar{\alpha}}\delta_{\rho\alpha'}-\delta_{\rho\bar{\alpha}}\delta_{\sigma\alpha'})\bar{n}_\beta n_{\alpha'}\kappa_\alpha,
\nonumber \\
\end{eqnarray}
the right-hand side of Eq. (\ref{K1}) becomes that of Eq. (\ref{K}). In a similar way it can be shown that the sum
$\sum_\mu(Y^\mu_{\lambda_1:\lambda_2\lambda_3})^*x^\mu_{{\alpha}}$ on the right-hand side of Eq. (\ref{yx}) becomes $K_{\alpha\lambda_2\lambda_3:\lambda_1}$
given by Eq. (\ref{K}).
The equations for $n_{\alpha\alpha'}$ and $\cal{C}_{\alpha\beta\alpha'\beta'}$ are also related to those
for $x^\mu_\alpha$, $y^\mu_\alpha$, $X^\mu_{\alpha\beta:\alpha'}$ and $Y^\mu_{\alpha':\alpha\beta}$.

\end{document}